\documentclass[twocolumn,floatfix,aps,prx]{revtex4-1}
\usepackage{epsfig,subfigure,amssymb,amscd,amsmath,graphicx}
\usepackage{hyperref}

\newcommand{\braket}[2]{\left\langle #1 | #2 \right\rangle}

\newcommand{\ket}[1]{\left|#1\right\rangle}

\newcommand{\elliptic}[3]{\vartheta_{#1}\!\left(\left.#2\vphantom{#3}\right|#3\right)}
\newcommand{\ellipticprime}[3]{\vartheta_{#1}^{\prime}\!\left(\left.#2\vphantom{#3}\right|#3\right)}
\newcommand{\ellipticgeneralized}[4]{\vartheta\left[\begin{array}{c} #1\\ #2\end{array}\right]\left(\left.#3\vphantom{#4}\right|#4\right)}

\newcommand{\ie}[0]{\emph{i.e.}\@ }
\newcommand{\eg}[0]{\emph{e.g.}\@ }
\newcommand{\Eg}[0]{\emph{E.g.}\@ }

\newcommand{\figref}[1]{FIG.~\ref{#1}}
\newcommand{\equref}[1]{Eq.~\eqref{#1}}
\newcommand{\refcite}[1]{Ref.~\onlinecite{#1}}
\newcommand{\refscite}[1]{Refs.~\onlinecite{#1}}

\newcommand{\picwidth}[0]{8.5cm}

\begin{document}

\title{Energy projection and modified Laughlin states}

\author{M.~Fremling$^{1,2}$, J.~Fulsebakke$^{2}$, N.~Moran$^{2}$, J.~K.~Slingerland$^{2,3,4}$}

\affiliation{
$^{1}\!\!\!$ Department of Physics, Stockholm University, SE-106 91 Stockholm, Sweden \\  
$^{2}\!\!\!$ Department   of  Mathematical  Physics,  National University of Ireland, Maynooth, Ireland.\\
$^{3}\!\!\!$ Dublin Institute for Advanced  Studies, School of Theoretical  Physics, 10 Burlington Rd, Dublin, Ireland.\\
$^{4}\!\!\!$ Rudolf Peierls Centre for Theoretical Physics, 1 Keble Road, Oxford OX1 3NP, United Kingdom.}

\begin{abstract}
\noindent
We develop a method to efficiently calculate trial wave functions for quantum Hall systems which involve projection onto the lowest Landau level.
The method essentially replaces lowest Landau level projection by projection onto the $M$ lowest eigenstates of a suitably chosen hamiltonian acting within the lowest Landau level.
The resulting ``energy projection'' is a controlled approximation to the exact lowest Landau level projection which improves with increasing $M$.
 It allows us to study projected trial wave functions for system sizes close to the maximal sizes that can be reached by exact diagonalization and can be straightforwardly applied in any geometry.
 As a first application and test case, we study a class of trial wave functions first proposed by Girvin and Jach\cite{Girvin84}, which are modifications of the Laughlin states involving a single real parameter.
 While these modified Laughlin states probably represent the same universality class exemplified by the Laughlin wave functions, we show by extensive numerical work for systems on the sphere and torus that they provide a significant improvement of the variational energy, overlap with the exact wave function and properties of the entanglement spectrum.
\end{abstract}

\maketitle

\section {INTRODUCTION}
\label{sec:Introduction}

Much of our understanding of the fractional quantum Hall effect (FQHE) and other strongly correlated systems comes from trial wave functions which describe the ground state and low lying excitations of the various phases of these systems.
A crucial step in constructing most of these trial wave functions is restriction of the Hilbert space of the system so that only single particle states in a low lying band can be occupied.
In the case of the FQHE, the single particle states are usually restricted to the lowest Landau level (LLL).
Sometimes it is possible to get a simple closed form expression for the trial states which satisfies this requirement,
as in the case of Laughlin's trial wave functions\cite{Laughlin83}, for the Hall conductance plateau at filling $\nu=\frac{1}{3}$.
However, more often, trial wave functions are constructed from some physical intuition without taking this restriction into account and then explicitly projected onto the LLL.
This is most famously the case for Jain's composite fermion (CF) wave functions\cite{Jain89,Jain_CF},
which give a good description of the physics of the Hall effect at fillings
throughout the region of the LLL where it is observed.
The simplest of these wave functions are of the form 
\begin{equation}
\label{eq:CF_basic}
P_{LLL}\left(\chi^{(CF)}(z_1,...,z_{N})\prod_{i<j}(z_{i}-z_{j})^{2p}\right).
\end{equation}
Here, $z_1,...,z_{N}$ are the complex coordinates of the $N$ electrons in the two-dimensional space.
The Jastrow factor $\prod_{i<j}(z_{i}-z_{j})^{2p}$ can be thought of as attaching $2p$ magnetic flux quanta to each of the electrons, or more naively, we can think of it as just lowering the correlation energy by keeping the electrons well separated.
The factor $\chi^{(CF)}$ is a Slater determinant built from single particle wave functions for a system at effective flux $N_{\Phi}^{(eff)}=N_{\Phi}- 2pN$.
This factor usually has nonzero occupation of states in higher Landau levels which means its polynomial part will depend on $\bar{z}_{i}$ as well as $z_{i}$.
The explicit orthogonal projection $P_{LLL}$ onto the LLL is then needed to bring the trial state back into the LLL Fock space.

The projection $P_{LLL}$ is hard to implement exactly.
The simplest method is to calculate the overlap of the unprojected trial states with the LLL Fock states by Monte Carlo integration, but this is only feasible for small system sizes.
Alternatively, one may use algebraic methods based on normal ordering and converting occurrences of $\bar{z}_{i}$ to derivatives $\partial_{z_{i}}$ (see \eg \refcite{Girvin84}), but this too can only be done for small system sizes.
In fact, considerable effort has been devoted to the development of approximate projection methods for these wave functions\cite{Jain97_PRB_55} and for the related CF wave functions with reverse flux attachment\cite{Moller_05_PRB,Davenport_2012_PRB}.
Only after the introduction of these methods has it been possible to probe even the largest system sizes which can be accessed by numerical diagonalization of the exact hamiltonian and then only for composite fermion type wave functions on a plane or sphere.
There is considerable interest now in studying systems on a torus, because this allows for a more direct examination of the topological order, for example through calculation of the Hall viscosity\cite{Fremling_14_PRB}.
Toroidal systems also lend themselves well to numerical study by density matrix renormalization group methods~\cite{Shibata01,Zaletel13}.
On a torus, until recently, there was not even a consensus on the correct form
of the CF trial wave functions, due to the fact that taking products of wave
functions, as in \eqref{eq:CF_basic}, does not satisfy the toroidal boundary conditions.
A number of recent works, \eg \refscite{Hermanns13_PRB,Suorsa11_PRB,Suorsa11_NewJPhys}, have introduced natural CF trial wave functions on the torus, but so far there is no efficient way of evaluating the required LLL-projection.
Looking beyond the CF paradigm, it is easy to write down a great many trial wave functions by employing  intuitive reasoning followed by explicit LLL-projection.
Many realizations of Haldane Halperin hierarchy wave functions \cite{Haldane83,Halperin84} which do not lie in the main CF series fall into this class, but much more is possible. A comprehensive overview of hierarchy constructions can be found in \refcite{Hansson16}.

The main aims of this paper are first of all, to introduce and test a projection method which will allow any proposed projected wave function to be studied up to the system sizes which can be reached by exact diagonalization of a reasonable local hamiltonian (\eg the Coulomb hamiltonian), as long as the real space form of the unprojected wave function can be easily evaluated.
We call this method the \emph{Energy Projection (EP)} and it simply consists of replacing the projection $P_{LLL}$ onto the lowest Landau level by projection onto a much lower dimensional space generated by low energy eigenstates of some hamiltonian.

Secondly, as a first application of this method, we study a set of trial wave functions which attempt to improve on Laughlin's wave function at filling $\nu=\frac{1}{q}$. These ``modified Laughlin states'' lower the correlation energy by inclusion of a factor which pushes the electrons further apart without changing the flux, or at least this is the naive intuition before projection.
Such wave functions were already proposed by Girvin and Jach in 1984\cite{Girvin84} and on the disk they take the form
\begin{equation}
\label{eq:alt-L_basic}
P_{LLL}\left(\prod_{i<j}(z_{i}-z_{j})^{q}\prod_{i<j}|z_{i}-z_{j}|^{2d} e^{-\frac{q+2d}{4q}\sum_k |z_{k}|^2}\right).
\end{equation}
When $d=1$, this can actually be interpreted as a state of composite fermions at
CF filling $\nu_{CF}=1$, with $q+1$ fluxes attached to each CF in the direction
opposite to that of the external field.
However, for other values of $d$, there is no such interpretation and the projection is not straightforward to perform even on the plane.
A torus version of these states can also be constructed and has been examined for up to $N=4$ particles in \refcite{Fremling13}. Using EP, we are able to study these wave functions in any geometry (we focus on sphere and torus here) and
at much larger sizes.

\section{Energy Projection}

In our projection scheme, we find a number of low energy eigenstates of some reasonable hamiltonian (most commonly the Coulomb hamiltonian) acting within the LLL.
We then project the trial wave function onto this set of low energy states.
The idea is that for any reasonable trial wave function, we will find nearly the entire projection onto the LLL using a number of eigenstates of the hamiltonian which is very small compared to the size of the LLL.
Before we analyze whether this approach really works, a rough analysis of the computation time involved is useful.
Particularly, we should compare the time it takes to calculate the energy
projection of a state to the time taken to calculate the exact projection. 

We will assume that the real space form of the trial wave function $\psi$ before LLL-projection can be easily calculated.
The exact LLL-projection of $\psi$ can then be found by calculating its overlaps $\braket{\phi_{\vec{n}}}{\psi}$ with the Fock states $\phi_{\vec{n}}$ labeled by the occupation numbers $\vec{n}$ of the LLL-orbitals.
Calculation of these overlaps can be done by Monte Carlo integration.
The number of orbitals in the LLL equals $N_{\Phi}$ (up to small geometry dependent corrections) and hence for a fermionic system, the number of Fock states spanning the LLL is $\binom{N_{\Phi}}{N}$.
The evaluation of a single Fock state (using Gaussian elimination to evaluate the determinant) scales as $N^3$, where $N$ is the number of particles.
If we denote by $N_{MC}$ the average number of evaluations per Fock state necessary to find the overlaps with $\psi$ to the desired accuracy, then the total time needed for the LLL-projection of $\psi$ scales as $N^3 N_{MC} \binom{N_{\Phi}}{N}$.
The factor $\binom{N_{\Phi}}{N}$ clearly increases very fast with both $N$ and $N_{\phi}$.
At fixed filling fraction $\nu$, we have $N_{\phi}\approx\nu^{-1}N$ and, defining $\xi=\nu^{-1}$ we see that  
\[\binom{N_{\Phi}}{N}\sim  \frac{1}{\sqrt{2 \pi N}} \sqrt{\frac{\xi}{\xi-1}}\left(\frac{\xi^{\xi}}{(\xi-1)^{(\xi-1)}}\right)^{N},\]
so the size of the Hilbert space increases exponentially in $N$. 
Unfortunately, $N_{MC}$ also tends to grow quickly with $N$.
Wave functions for strongly interacting systems tend to have nonzero overlaps of comparable size with a significant fraction of the Fock states in the LLL.
Hence, if we define $f=|\braket{\psi}{P_{LLL}\psi}|^2$, then the Monte Carlo integration has to resolve overlaps of typical size $\sqrt{f/\binom{N_{\Phi}}{N}}$.
This means that the allowable error on a given overlap should also be very small.
In fact, if the typical error on the overlaps is $\epsilon$, then if we are very optimistic and take the errors on the overlaps to be independent of each other, we expect an overall error on the projection of $\psi$ which is of order $\sqrt{\binom{N_{\Phi}}{N}}\epsilon$.
To bring this back to something of order $\sqrt{f}$, we require that $\epsilon\sim \sqrt{f/\binom{N_{\Phi}}{N}}$.
The statistical error in the Monte Carlo integration will normally be inversely proportional to the square root of the number of independent Monte Carlo samples generated, and we expect that the number of independent samples will be (at most) of order $\frac{N_{MC}}{N}$.
Hence for the desired accuracy, we require that $N_{MC}\sim N\binom{N_{\Phi}}{N}/f$.
This yields an (optimistic) estimate of the scaling of computation effort for exact LLL-projection as $f^{-1}N^4 \binom{N_{\Phi}}{N}^2$. 

Clearly, this is problematic in studying large systems.
Nevertheless naively, it would seem that obtaining exact states for comparison to these trial states may be even more onerous, as numerical diagonalization scales naively as the third power the Hilbert space dimension, \ie $\binom{N_{\Phi}}{N}^3$.
In practice however, the hamiltonians of interest are relatively sparse, usually have a high degree of symmetry and we are normally only interested in a small number of low lying eigenstates.
As a result, the LLL-projection of the trial wave functions as described here nearly always becomes impossible at system sizes considerably smaller than those accessible to exact diagonalization methods. 

Now consider the computational effort needed to perform energy projection.
We will assume first of all that we can get a good approximation of the exact
projection using the lowest $M$ eigenstates of the chosen hamiltonian.
Here $M$ should be a number that does not grow quickly with $N$, in particular, $M\ll \binom{N_{\Phi}}{N}$.
We will ignore the computational effort needed to obtain the $M$ lowest eigenstates of the hamiltonian -- we will usually work at system sizes where this is not the bottleneck of the computation.
The energy projection is simply the projection onto the $M$-dimensional
subspace of the Hilbert space spanned by the $M$ lowest eigenstates.
It involves calculating the overlaps $\braket{i}{\psi}$, where $\ket{i}$, $i\in\{1,...M\}$ labels the lowest $M$ eigenstates of the hamiltonian.
Each state $\ket{i}$ involves up to $\binom{N_{\Phi}}{N}$ Fock states. Hence a single evaluation of the state $\ket{i}$ in real space takes effort of order $N^3\binom{N_{\Phi}}{N}$.
Once the Fock states have all been evaluated, we can keep their values (subject to memory constraints) and use them to evaluate the other $M-1$ eigenstates of the hamiltonian.
Evaluation of the $M$ overlaps needed for the energy projection then takes effort of order $(N^3 +M-1)\binom{N_{\Phi}}{N} N'_{MC}$, where $N'_{MC}$ is the average number of evaluations of the states $\ket{i}$ needed to get good Monte Carlo estimates of the overlaps.
The main advantage of energy projection is that the individual overlaps involved should now be much larger than the overlaps with individual Fock states.
We now expect the average overlap to be $\sqrt{f/M}$ and to get the error on the projection of $\psi$ to be of order $\sqrt{f}$, we will need $N'_{MC}\sim N M/f$, giving an estimate for the total computational effort involved in energy projection as $f^{-1}M N (N^3 + M-1) \binom{N_{\Phi}}{N}$.
This is clearly much better scaling than exact projection, as long as $M\ll  \binom{N_{\Phi}}{N}$ and $M<N^3$ which usually the case, or in any event if  $M^2\ll  \binom{N_{\Phi}}{N}$. 
This scaling can be further improved if the trial wave function is an eigenstate of some symmetry of the hamiltonian, because in that case, we only need to consider overlaps with eigenstates with the same symmetry, reducing $M$.  
It turns out that energy projection usually allows us to work with trial wave functions at system sizes  close to the largest sizes accessible to exact diagonalization. 
Of course, it must be remembered that energy projection is an approximation, as we are throwing away components of the trial wave function at higher energies.
For reasonable trial wave functions we can hope that these components are small.
In the next sections we will investigate in some detail whether this is actually the case. 

\section{Testing the projection}

\begin{figure*}[htb]
 \begin{center}
    \begin{tabular}{cc}
      Sphere & Torus\\
	\includegraphics[width= \picwidth]{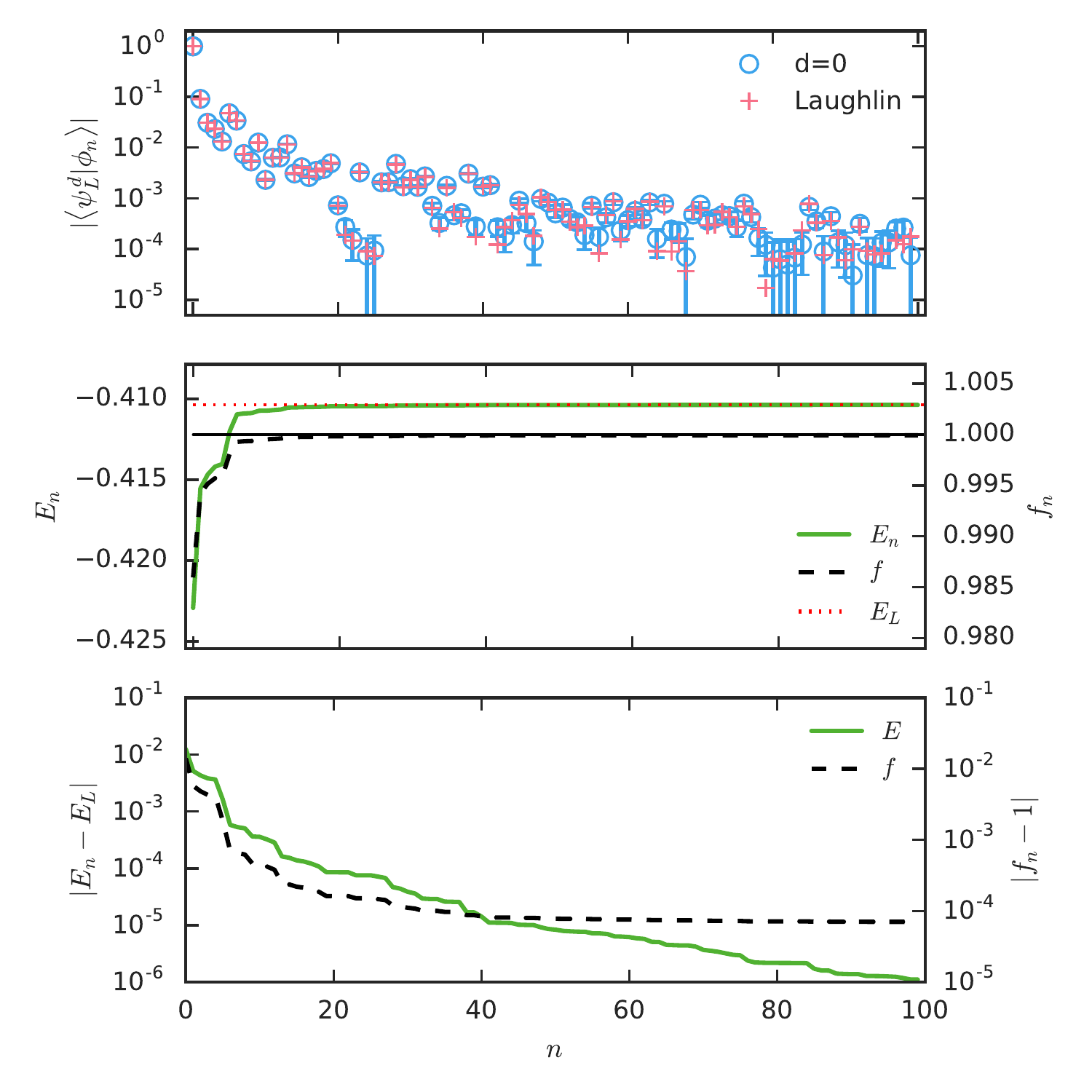} 
	&
	\includegraphics[width= \picwidth]{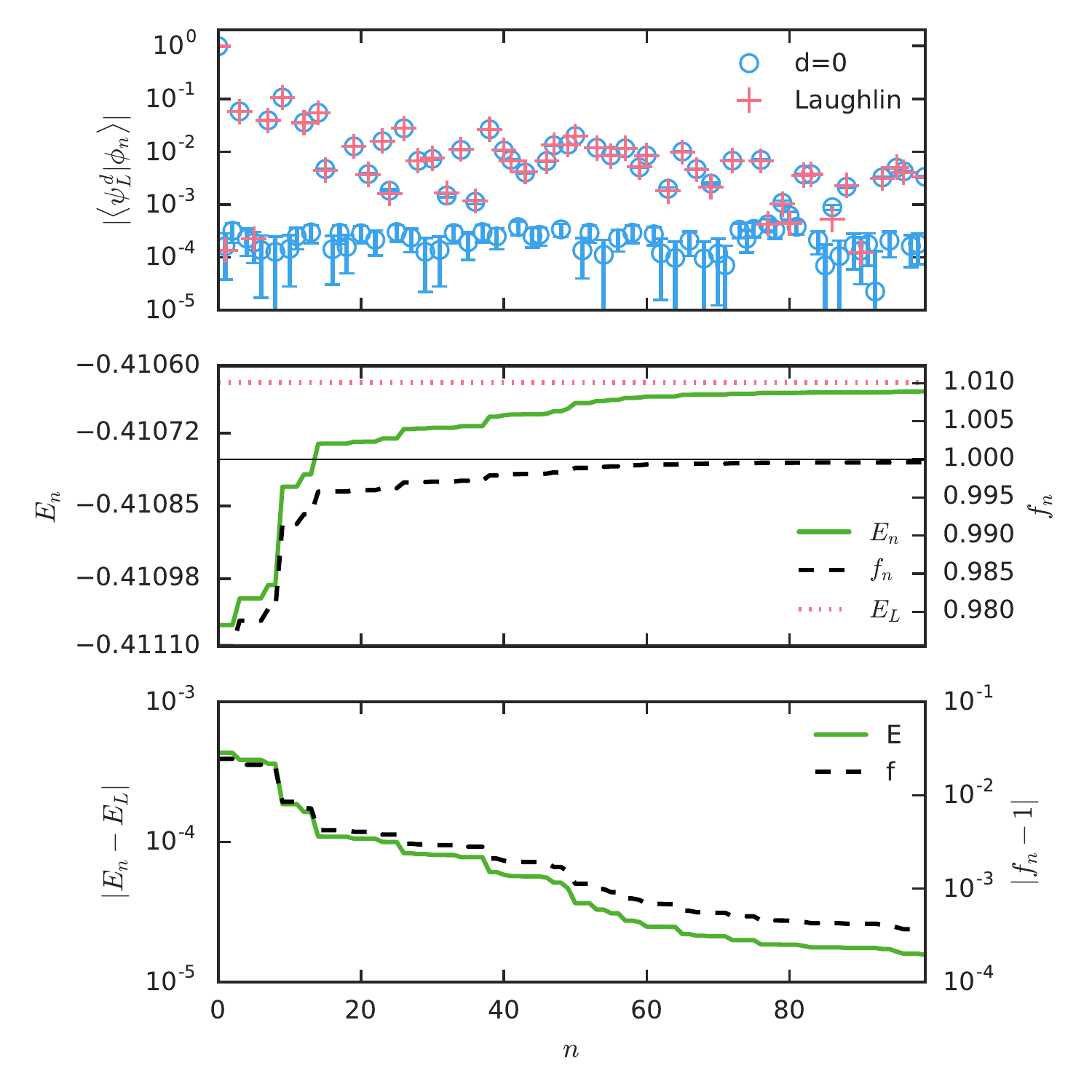}       
     \end{tabular}
\caption{{\small 
{\bf Left Panels (Sphere):} Overlaps and variational energies for $N=10$
electrons on a sphere (full Hilbert space dimension  is of order $10^7$, after
use of symmetries 319), using $1.2\times 10^8$ MC samples. Eigenstates
calculated using iterative diagonalisation methods discussed in section
\ref{sec:computations}. \\
\emph{Upper:} Exact and approximate (indicated by $d=0$) overlaps of the $ \nu = 1/3$ Laughlin wave function with consecutive LLL-Coulomb eigenstates (with the Coulomb ground state on the left). \\
\emph{Middle:} Cumulative square overlap $f$ and energy estimate. The expected limit values for $f$ and $E$ are indicated. \\
\emph{Lower:} Absolute differences between $E_n$ and $f_n$ (note the logarithmic scale).
\\
{\bf Right Panels (Torus):} Overlaps and variational energies for $N=10$ electrons
on a square torus (full Hilbert space dimension is of order $10^7$, after use of
symmetries of order $5 \times 10^5$), using $2\times 10^7$ MC samples.
Panel content is the same as for panels on the left hand side. The appearance of many near zero overlaps in the upper panel is due to the $C_4$ symmetry of the system on the square torus. 
}}
\label{fig:d_0_Ef}
\end{center}
\end{figure*}

To establish how well the energy projection method works we perform a number of tests.
A first test would be to select some trial wave function,
calculate the exact projection using Monte Carlo evaluation of overlaps with the Fock basis and compare it to the energy projection.
However, for system sizes which are small enough to allow for accurate calculation of the exact projection in this way,
we can also calculate the full spectrum of the Coulomb hamiltonian in the LLL and use energy projection using the full spectrum,
which is in effect also exact projection and moreover with a smaller error than the projection using the Fock basis.
Therefore we have tested the energy projection first of all on trial wave functions which are fully in the LLL.
This means the projection is redundant, but allows us to see how well the energy projection reproduces the full state.
Here we present results for the $\nu=1/3$ Laughlin wave-function $\psi_{L}$, on the sphere and torus, energy projected using the Coulomb hamiltonian, see \figref{fig:d_0_Ef}. 

The upper panels of \figref{fig:d_0_Ef} show the exact and approximate overlaps of
$\psi_L$ with the $n$:th energy eigenstate of the Coulomb hamiltonian.
The exact wave function $\psi_{L}$ was computed by diagonalizing the short range hamiltonian based on Haldane's pseudopotentials\cite{Haldane83} for which it is the unique ground state.
Since this give us $\psi_{L}$ in the same basis as the Coulomb eigenstates, it is easy to obtain numerically exact overlaps also.
The approximate overlaps are marked by $d=0$, in reference to the later use of nonzero values of $d$ when we modify the Laughlin wave function.
These overlaps were obtained directly by performing Monte Carlo integration in real space.
We find that on both sphere and torus all overlaps larger than $10^{-3}$ can be well resolved by the MC estimates.
This can obviously be improved by taking more Monte Carlo samples.
In the torus plot we see many small overlaps at a level just above $10^{-4}$.
Most of these overlaps are actually exactly zero;
the system on a square torus has a $C_4$ symmetry and hence the eigenstates of the hamiltonian with symmetry behavior different from $\psi_{L}$ have zero overlaps.
The nonzero values observed give us a useful idea of the accuracy of the Monte Carlo overlaps. 
Note that -- as expected -- the (nonzero) overlaps are diminishing as a function of $n$, and the declining trend is clearly visible even on the logarithmic scale.
This confirms the physical intuition that most of the projected state is captured by the low energy excitations and higher energy excitations become less and less important. 

The middle panels show the cumulative square overlap
\[f_n=\sum_{j=1}^n |c_j|^2, \]
and the reconstructed energy 
\[ E_n=\frac{1}{f_n}\sum_{j=1}^n \epsilon_j |c_j|^2. \] 
Here $c_j=\braket{\phi_j}{\psi}$ is the overlap between energy eigenstate no. $j$ and the unprojected wave function $\psi$, and $\epsilon_j$ is the energy of that $j$:th eigenstate. 
When $n$ approaches the total number of states in the LLL, $f_{n}$ represents the LLL content of the state (in this case we know this equals $1$) and $E_{n}$ becomes the variational energy of the state.
For comparison and guidance, the exact the Coulomb energy of the Laughlin state, $E_L$ and the limit value $f=1$ are included.
Since $f_n$ measures the $LLL$ content, it increases monotonically with $n$ and
saturates quickly for $n \ll N$ (note that these panels do not use a logarithmic scale).
As the energy levels are ordered ( $\epsilon_i>\epsilon_j$ for $i>j$ ) and $|c_j|\geq 0$, $E_n$ is also monotonic, and we note again that $E_n$ converges fast as a function of $n$.

In the lower panels we show the differences $|E_n-E_L|$ and $|f_n-1|$ on a logarithmic scale.
We see that 99.9\% of $\psi_L$ is already captured by using $n\approx50$ states on the torus and $n\approx15$ states on the sphere.
The energy is reproduced to within four decimals by taking $n\approx 20$ states on the torus and $n\approx30$ states on the sphere.
The stepwise behavior of these graphs is explained by considering their dependence on $c_{i}$. For example, the difference between two consecutive energy estimates is $E_{n+1}-E_n=\frac{|c_{n+1}|^2}{f_{n+1}}(\epsilon_{n+1}-E_n)\geq0$ and will thus jump when  $|c_{n+1}|^2$ is large, which is at the same time that $f$ jumps.
\begin{figure*}[htb]
  \begin{center}
    \begin{tabular}{cc}
      Sphere, $d=0$ & Torus, $d=0$ \\
        \includegraphics[width= \picwidth]{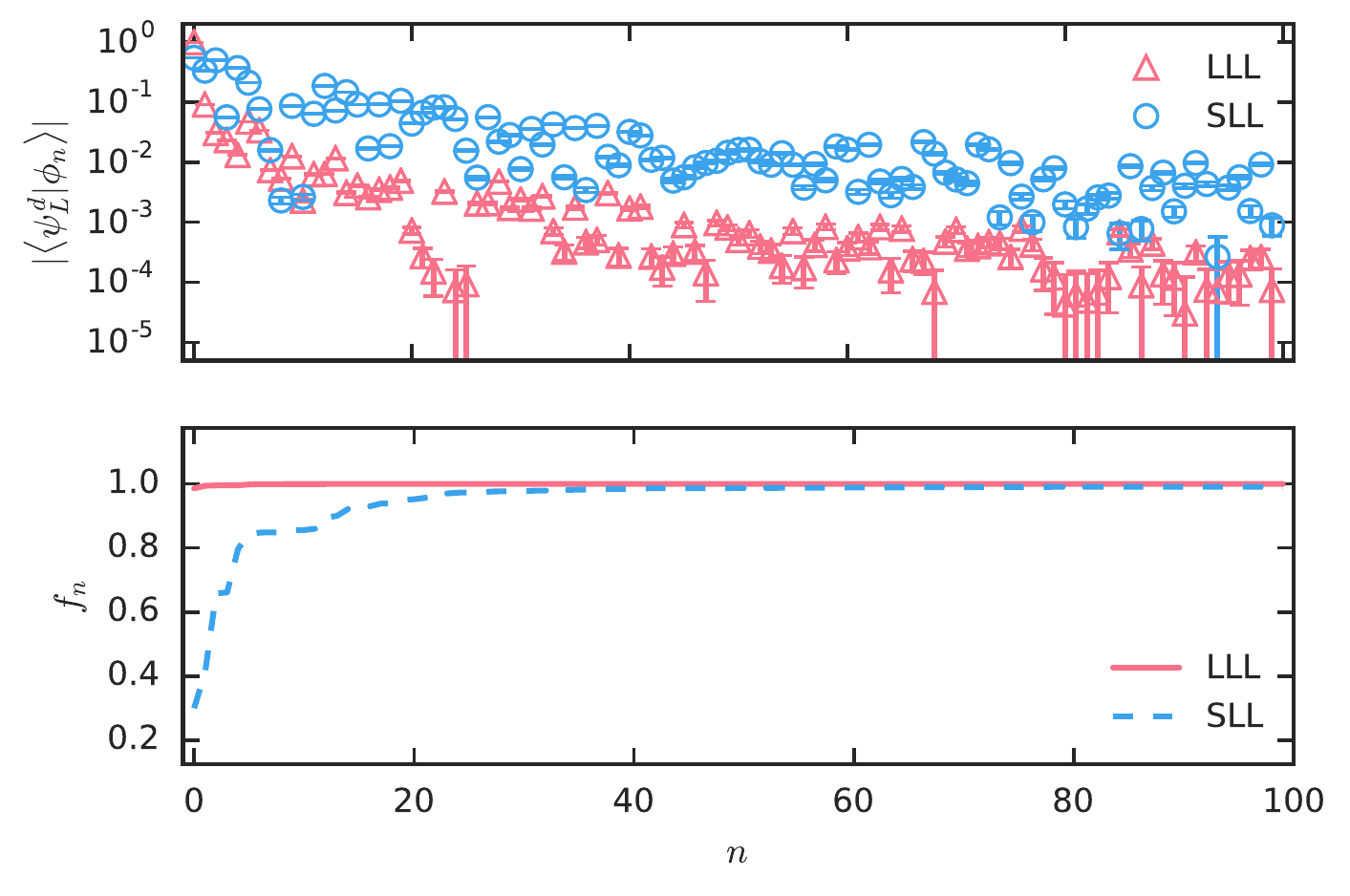} &
        \includegraphics[width= \picwidth]{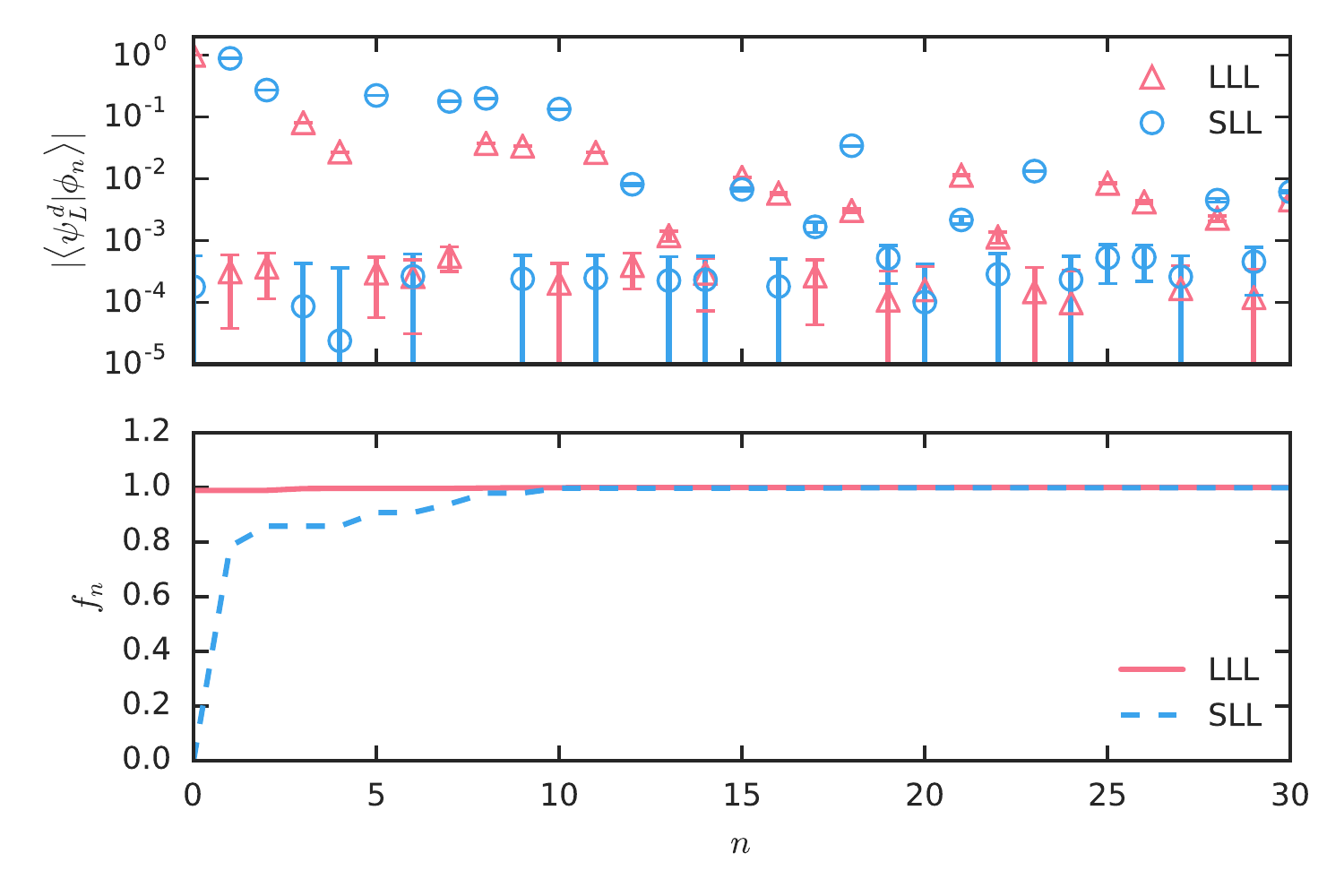} \\
       \hphantom{Sphere,} $d=1$ & \hphantom{Torus,} $d=1$ \\
        \includegraphics[width= \picwidth]{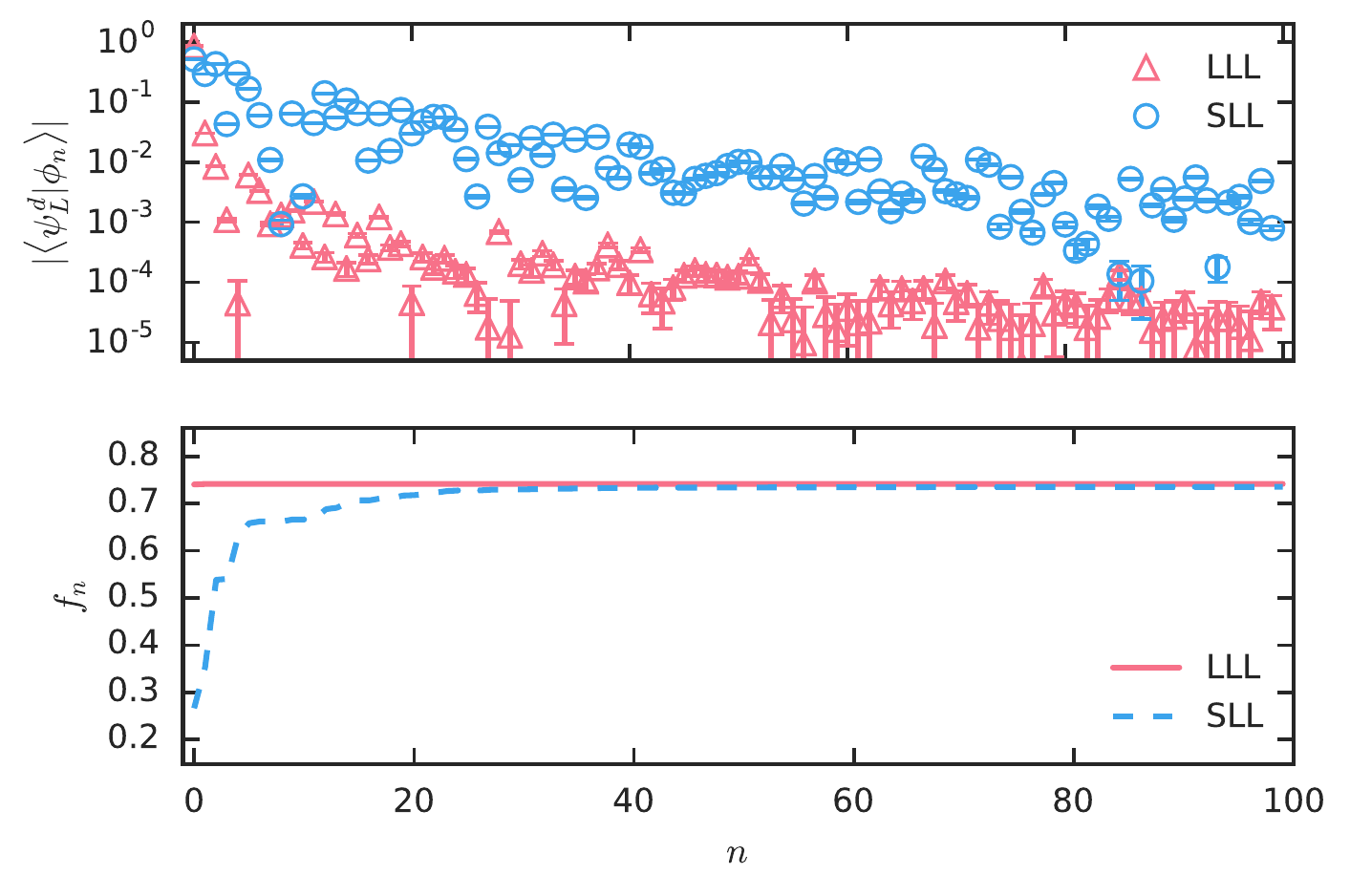}&
        \includegraphics[width= \picwidth]{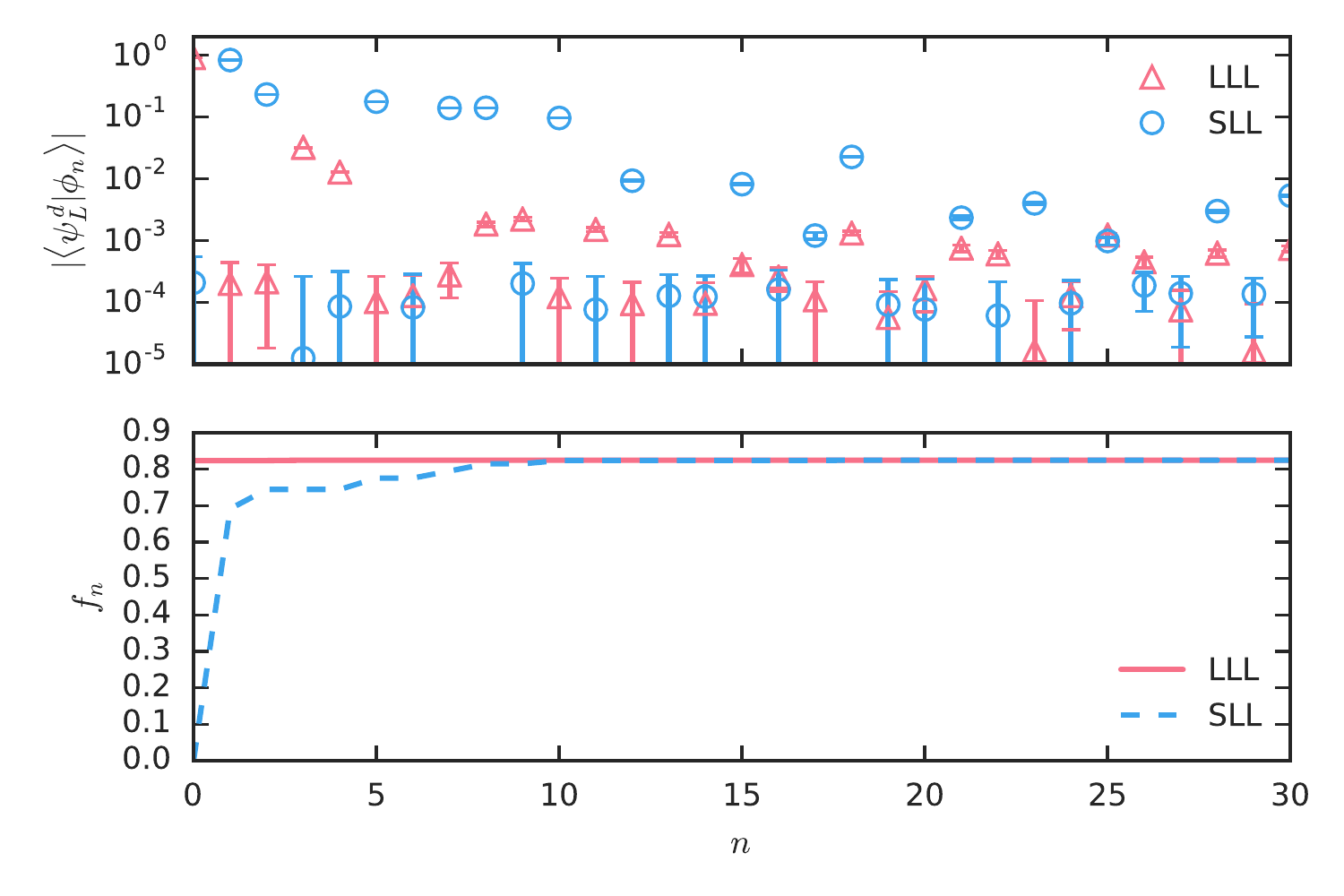} \\
       \hphantom{Sphere,} $d=7$ & \hphantom{Torus,} $d=6$ \\
        \includegraphics[width= \picwidth]{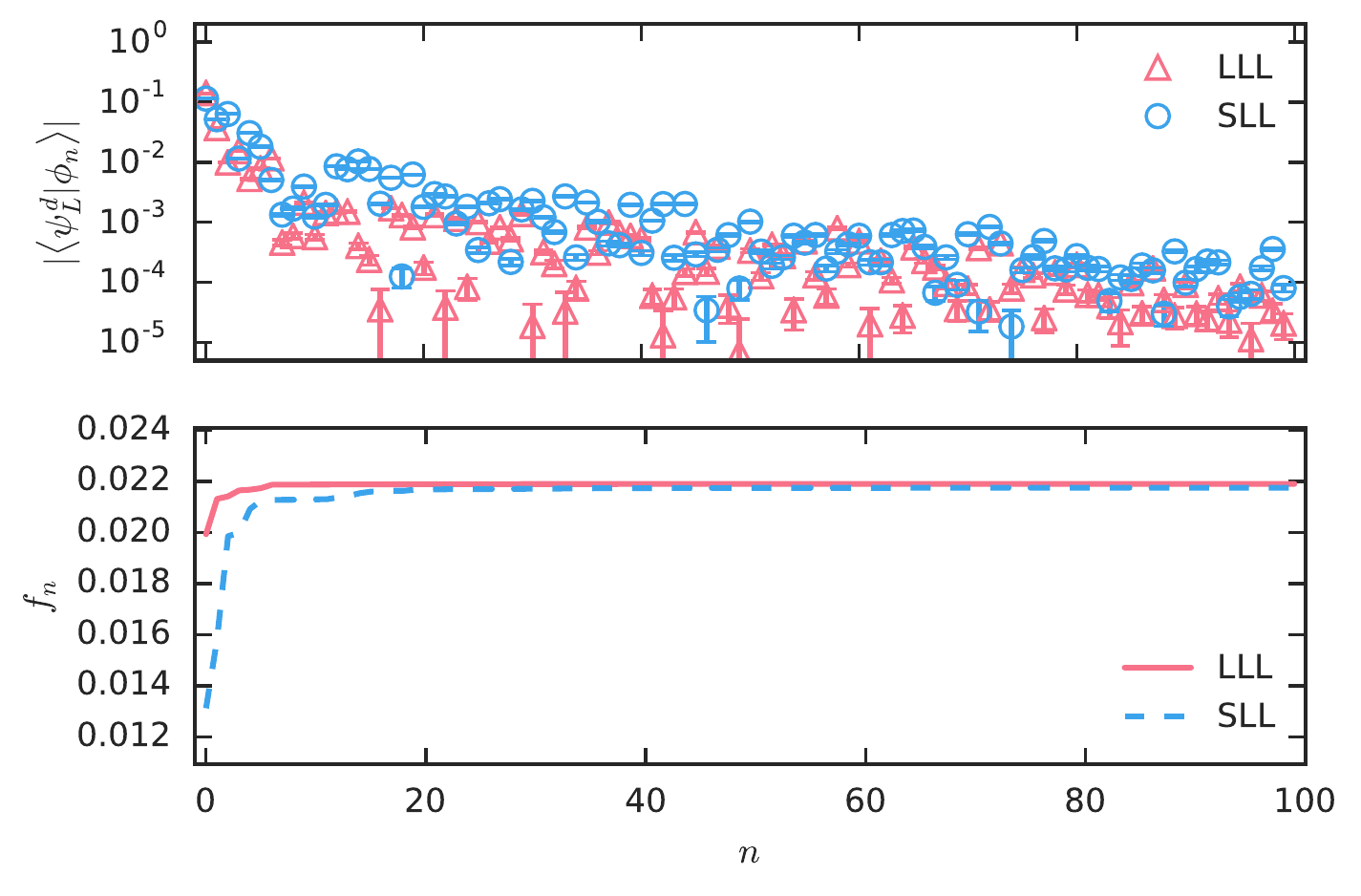}&
        \includegraphics[width= \picwidth]{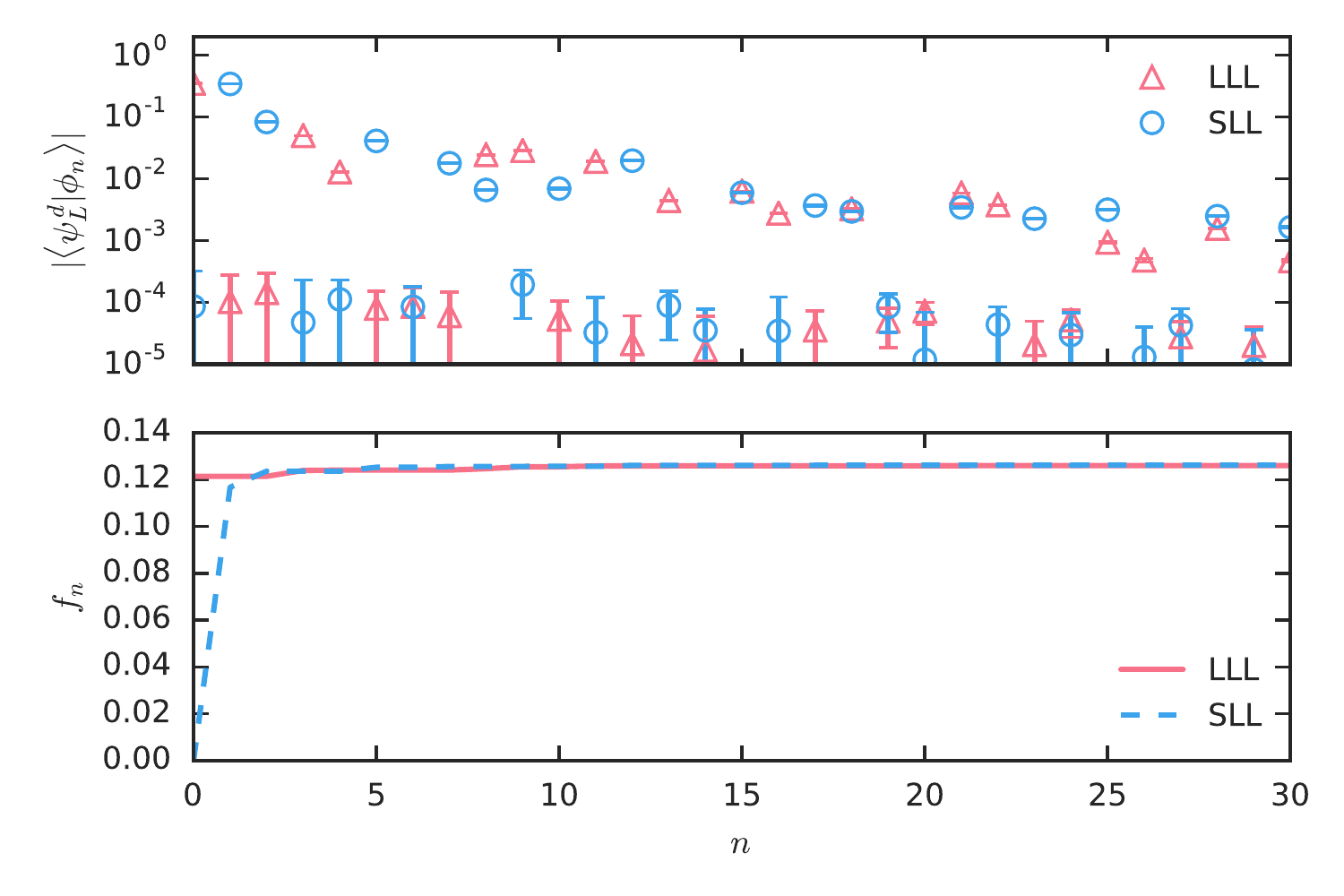}\\ 
    \end{tabular}
    \caption{{\small 
        Overlaps (upper panels in each figure) and cumulative squared overlaps
        (lower panels) with eigenstates of the lowest Landau level (LLL) and
        second Landau level (SLL) Coulomb potentials. Eigenstates calculated using
        iterative diagonalisation methods discussed in section \ref{sec:computations}. \\
\textbf{Left panels (Sphere):} Results for the modified Laughlin states on a sphere at filling $\nu=\frac{1}{3}$ for $d=0$ (top), $d=1$ (middle), $d=7$ (lower), for $N=10$ electrons.\\
\textbf{Right panels (Torus):} Results for the modified Laughlin states on a square torus at filling $\nu=\frac{1}{3}$ for $d=0$ (top), $d=1$ (middle), $d=6$ (lower), for $N=6$ electrons.\\
}}
    \label{fig:LLL_v_SLL_torus_d_0126_f}
  \end{center}
\end{figure*}

\begin{figure*}[htb]
  \begin{center}
    \begin{tabular}{cc}
      Sphere & Torus\\
      \includegraphics[width= 8.5cm]{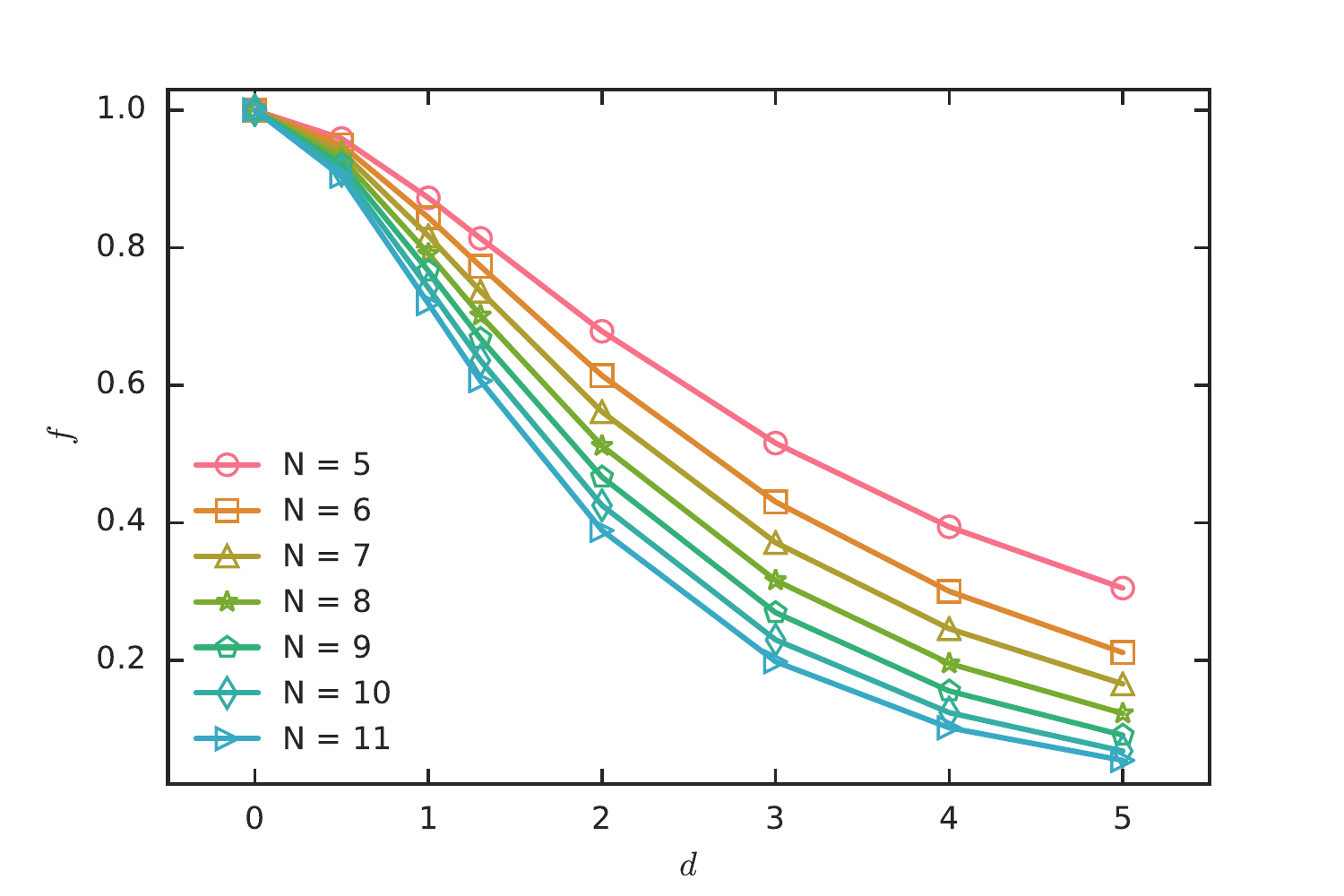}&
      \includegraphics[width= 8.1cm]{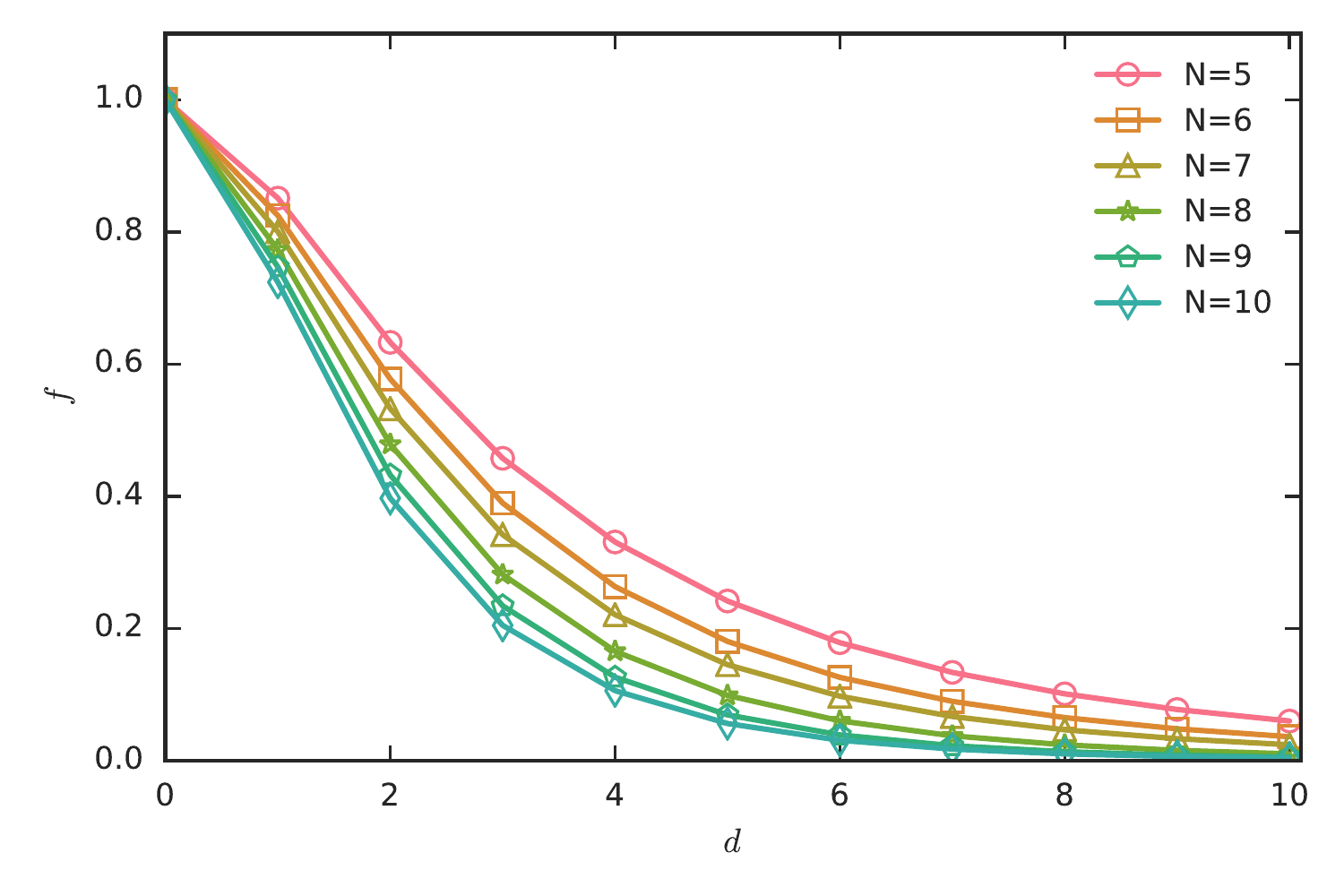}
    \end{tabular}
    \caption{\small 
      LLL--content $f$ of the modified Laughlin state as a function of $d$ and $N$ on the sphere (left) and torus (right)}
    \label{fig:LLL_content_vs_d_N}
  \end{center}
\end{figure*}

Next we test whether the energy projection is stable against changing the hamiltonian.
For the energy projection to be generically useful, its success should not depend crucially on which hamiltonian is used.
While the low energy sector of the hamiltonian should capture the state that is being projected, the detailed structure of the low energy states should not be important.
 We would expect that hamiltonians with completely different ground states should be viable, as long as they incorporate \eg repulsive interactions between the particles.    

Here, we compare the energy projection using the lowest LL and second LL Coulomb hamiltonians obtained by extracting pseudopotentials a la Haldane\cite{Haldane83} from the real space Coulomb hamiltonian using the lowest and second LL orbital wave functions.
 It is well known that these hamiltonians provide a completely different set of ground states at most accessible values of the flux and electron number.
 In particular, $\psi_{L}$ is an excellent trial wave function for the ground state of the LLL but not for the second LL Coulomb hamiltonian, where it has squared overlap of order at most $0.4$ with the ground state for systems of up to $15$ particles\cite{Ambrumenil88,Balram13}.
  Significant efforts have recently been made to determine whether the ground state of the SLL Coulomb hamiltonian even represents the same topological order as the Laughlin wave function\cite{Zaletel15,Johri14,Balram13} and there is also recent work on alternative wave functions which may improve the overlap\cite{Jeong16}. 
In addition to $\psi_{L}$ we show the $d=1$, $d=2$ and $d=6$ (torus) and $d=7$ (sphere) modified Laughlin states introduced in \equref{eq:alt-L_basic}.
Explicit expressions for the modified wave functions on the sphere and torus are given in \eqref{eq:Mod_Laughlin_Sphere} and \equref{eq:Mod_Laughlin_Torus}.
 Note that these $d\neq0$ wave functions are not entirely contained within the LLL so the energy projection is not redundant for these.

In \figref{fig:LLL_v_SLL_torus_d_0126_f} we show results for $N_e=6$ particles
on the torus (right panels, the small system size is chosen for illustrative
purposes) and for $N_e=10$ particles on a sphere (left panels). 
The upper panels in each subfigure again show the overlap for the $n$:th eigenstates and the lower panels show the cumulative content $f$ for the LLL (red) and SLL (blue) hamiltonians.
 Looking first at $d=0$ (or $\psi_L$, upper panels) we see that, just as in \figref{fig:d_0_Ef}, the LLL overlap falls off rapidly with increasing $n$ and that $f$ converges to good precision with only a few terms.
Comparing this to the SLL hamiltonian (blue), we see that the SLL ground state
and $\psi_L$ have a small overlap (zero within error on the torus and squared overlap of less than $0.4$ on the sphere).
On the other hand, practically all of $\psi_L$ is still captured by the low energy states.
In the toroidal system, after including as little as $n=15$ states (out of a
total of $1038$ states with the same total momentum), projection using the LLL and SLL hamiltonians both give $f=1$ to within $10^{-3}$.
 The spherical system also clearly gives projections from the two hamiltonians which are in close agreement, although at this system size more eigenstates of the SLL hamiltonian are needed. 

Turning our attention to $d=1$ (middle panels), we see qualitatively the same behavior.
However -- as for all $d\neq0$ -- the modified Laughlin wave functions are not contained within the LLL anymore,
so $f<1$ and we find limiting values of $f\approx0.8$ (torus) and $f\approx0.74$ (sphere). 
Nevertheless, it is clear from the figures that the energy projection still works, both with the LLL hamiltonian and with the SLL hamiltonian, as the $c_i$ decrease with $n$ and the low lying $c_i$ are still large enough to be accurately determined.
If the LLL content of the unprojected wave function is very small, then the accuracy will also be reduced as the smaller $c_j$ incur larger  relative errors in the MC-projection.
 Nevertheless, it is still possible to extract perfectly viable energy projections for considerably higher values of $d$. \Eg the lower left panel shows the $d=6$ system on the torus, where $f$ is only about $0.12$ and the lower right panel shows the $d=7$ system on the sphere, with $f\approx0.022$.
Both of these panels show that the SLL hamiltonian is more competitive with the LLL hamiltonian at larger values of $d$.
For the $d=6$ torus state, the SLL hamiltonian actually manages to capture the LLL content faster than the LLL hamiltonian at larger $n$.
In fact $f^{(\mathrm{SLL})}_n\gtrsim f^{(\mathrm{LLL})}_n$ for $3<n<10$, which shows that it is not always best to have good overlap with the ground state,
since it may sacrifice weight in the other low energy states and lead to a lower value of the total weight $f$.
 For the $d=7$ system on the sphere, the LLL hamiltonian wins out over the SLL hamiltonian throughout,
 but we can nevertheless observe that the overlaps are considerably closer than at $d=0$ or $d=1$. 
\begin{figure*}[htb]
\begin{center}
  \begin{tabular}{cc}
      Sphere & Torus\\
    \includegraphics[width= \picwidth]{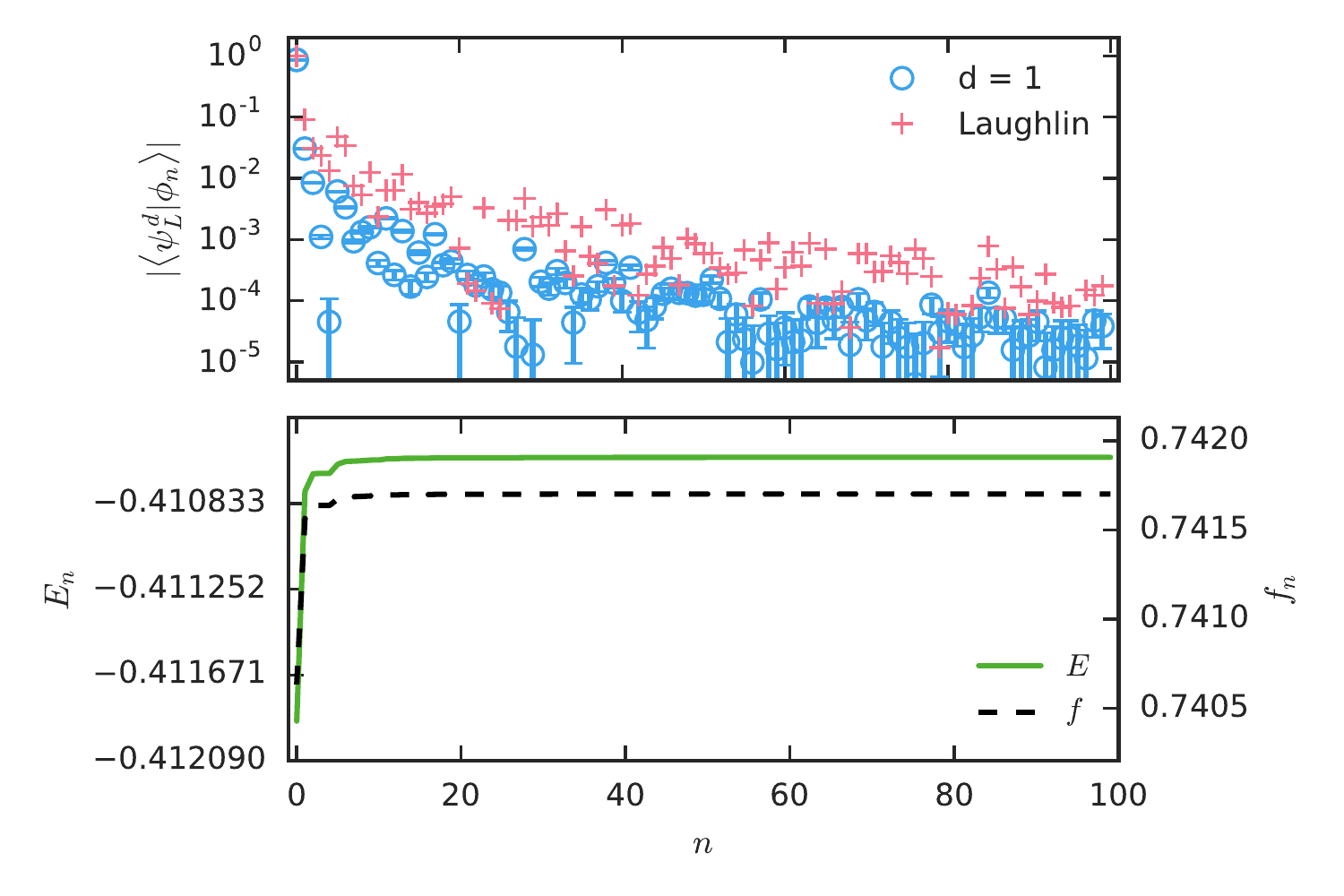}
    & 
    \includegraphics[width= \picwidth]{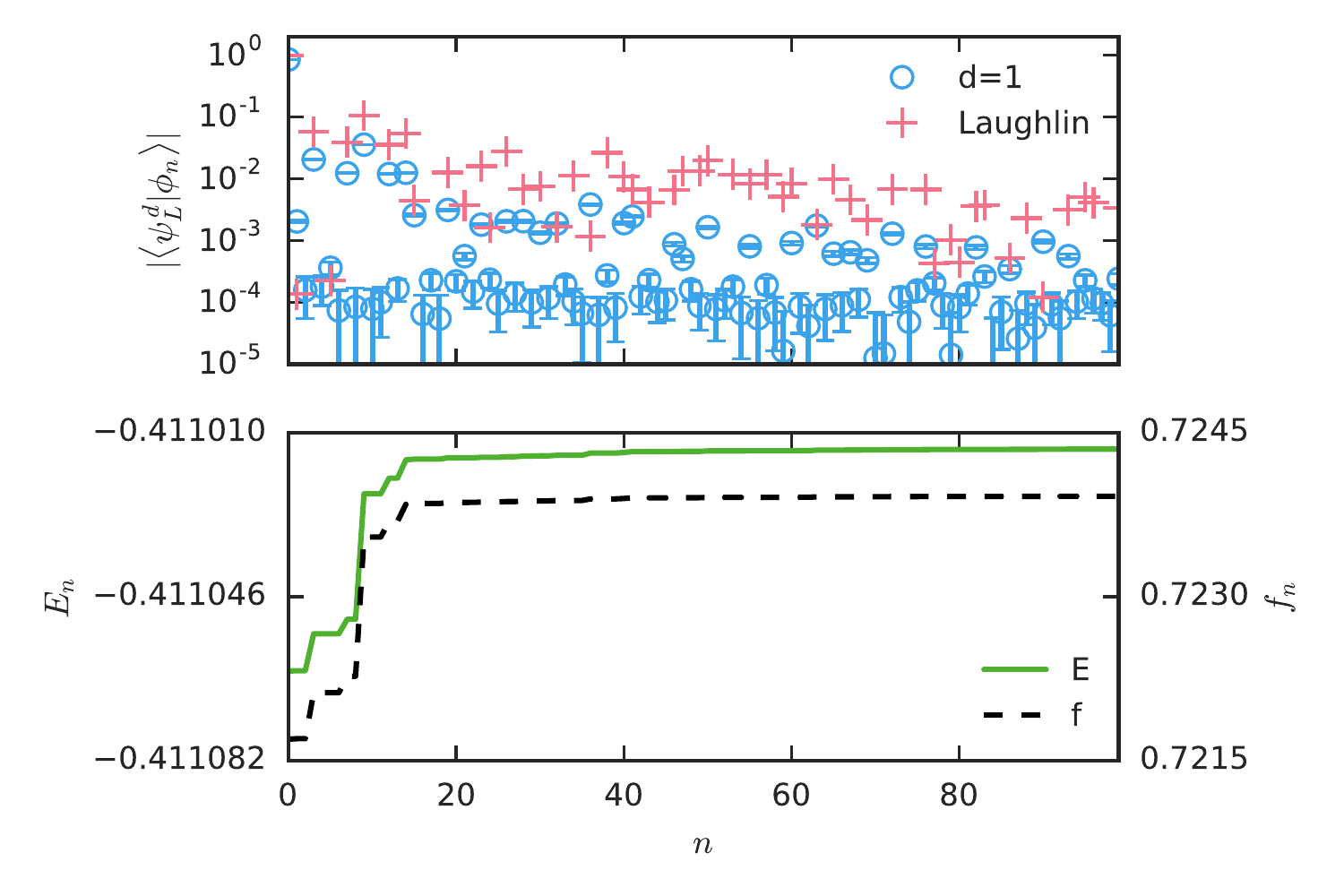}
    \\
    \includegraphics[width= \picwidth]{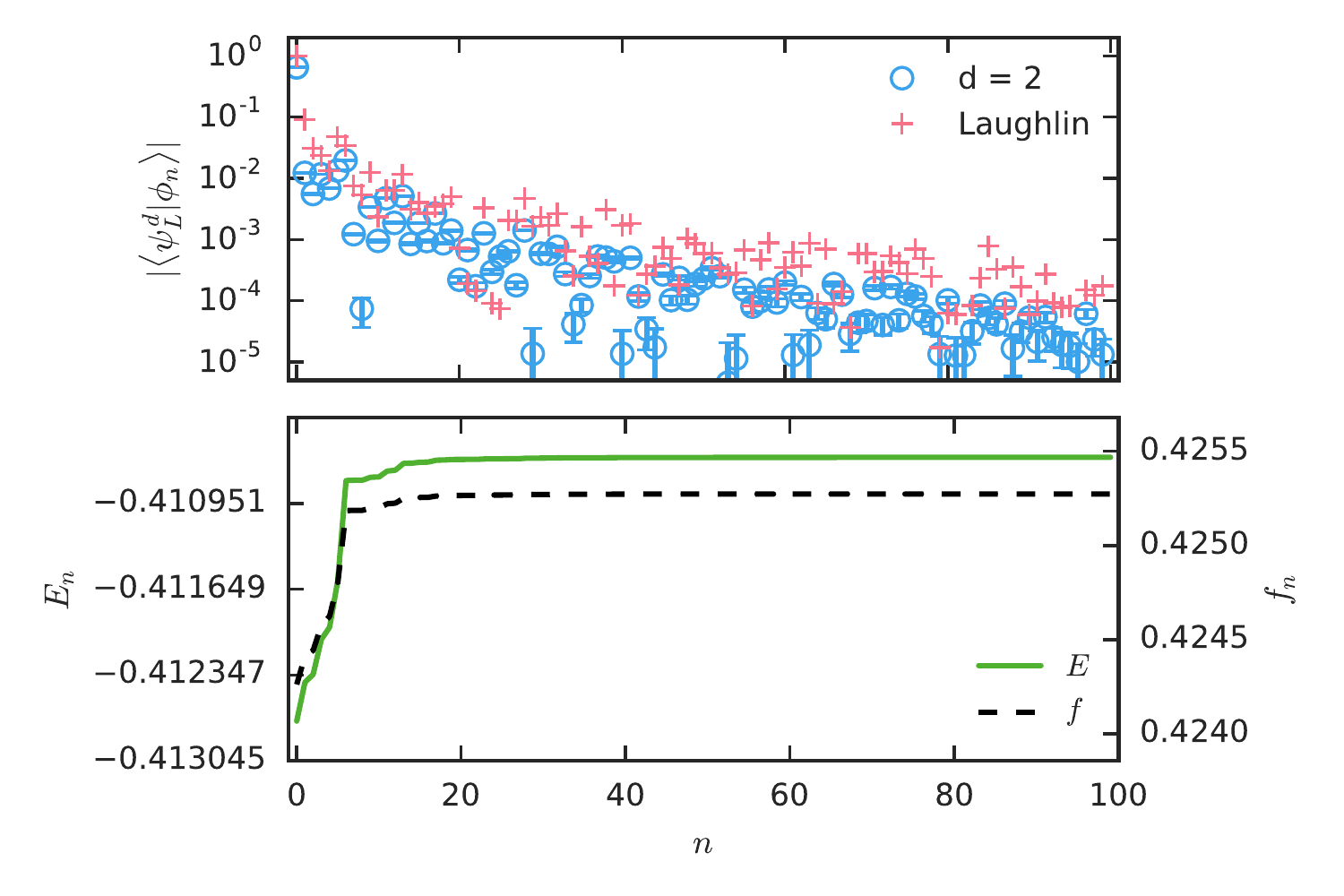}
    &
    \includegraphics[width= \picwidth]{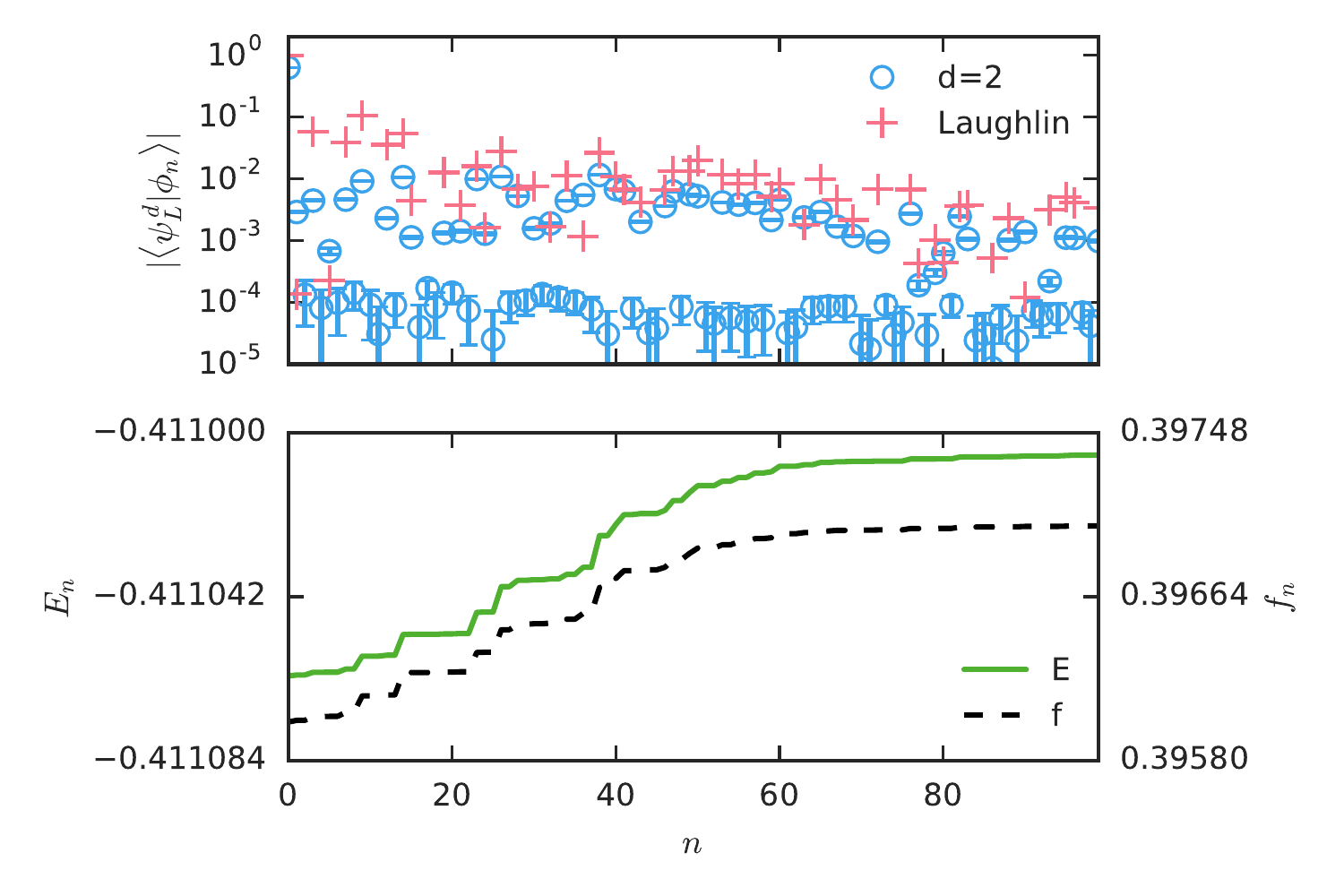} 
  \end{tabular}
  \caption{\small 
    Overlaps (upper panel in each figure) and cumulative squared overlaps (lower
panels) with eigenstates of the lowest Landau level (LLL) Coulomb
potentials. Eigenstates calculated using iterative diagonalisation methods
discussed in section \ref{sec:computations}. \\
    \textbf{Left panels (Sphere):} Results for modified Laughlin states at filling
    $\nu=\frac{1}{3}$ for $N=10$
    particles on the sphere with $d=1$ (top) and $d=2$ (bottom).\\
    \textbf{Right panels (Torus):} Results for modified Laughlin states on a square
    torus at filling $\nu=\frac{1}{3}$ for $d=1$ (top) and $d=2$ (bottom) for $N=10$ electrons.\\
  }
  \label{fig:d_1_d_2_Ef}
\end{center}
\end{figure*}

All in all \figref{fig:LLL_v_SLL_torus_d_0126_f} shows that the energy projection method is stable to appreciable changes of the projecting hamiltonian as long as the low energy content is preserved and the state has reasonable weight in the LLL.
We can also comment briefly on the suitability of the modified Laughlin wave functions as improved trial wave functions for the $\nu=7/3$ quantum Hall plateau.
 We find that the overlap with the ground state of the $SLL$ Coulomb hamiltonian on the sphere does improve when $d>0$ (as compared to $d=0$), but the improvement is not spectacular.
 The highest overlap we obtained was $0.603(2)$ for a system of $N=10$ particles on a sphere at $d=7$ (shown in \figref{fig:LLL_v_SLL_torus_d_0126_f}).
 On the torus the overlap is always found to be zero within error, which appears to signal that  the ground state of the SLL hamiltonian is in a different $C_4$ symmetry sector from that of the LLL hamiltonian.

Next we examine the speed of convergence for modified Laughlin states, as well as the LLL--content $f$ of these states, using the LLL Coulomb hamiltonian. 
Results for the limiting value of $f$ as a function of $d$ and $N$ are shown in figure~\figref{fig:LLL_content_vs_d_N}. 
As expected, the value of $f$ decreases both with increasing $N$ and with increasing $d$.
 However, this decrease is perhaps not as fast as might be naively expected, with appreciable LLL--content still remaining even at the largest system sizes probed, especially at low values of $d$.
 Results on the convergence of $f$ and of the variational energy for $d=1$ and $d=2$ with $N=10$ electrons are shown in \figref{fig:d_1_d_2_Ef}. 
On the sphere (left panels), both $f$ and the energy stabilize very quickly at $f\approx0.74$ for $d=1$ and $f\approx0.425$ for $d=2$.
Similarly, on the torus, we find $f\approx0.72$ for $d=1$ and $f\approx0.397$
for $d=2$. Table \ref{tab:convergence} shows how fast the energy estimate
and cumulative overlap converge for $d=1$ and $d=2$. It is clear from the table
that in both cases these values converge rapidly, but that for $d=1$, the
convergence is faster.

\begin{center}
\begin{table}
\subtable[\ Sphere]{
\begin{tabular}{ c |c ||  c |c |c |c |c }
  &m  & 3 & 4 & 5 & 6 & 7 \\
  \hline
  $d=1$ & $n^E$ &  2 & 3 & 12 & 33 & 67  \\
  \cline{2-7}
        & $n^f$  & 2 & 3 & 12 & 29 & 58  \\
  \hline
  $d=2$ & $n^E$ & 7 & 14 & 29 & 52 & 79  \\
  \cline{2-7}
        & $n^f$  & 2 & 7 & 19 & 42 &  74
\end{tabular}
\label{tab:sphere_convergence}
}
\subtable[\ Torus]{
\begin{tabular}{ c |c ||  c |c |c |c |c }
  &m  & 3 & 4 & 5 & 6 & 7 \\
  \hline
  $d=1$ & $n^E$ &   & 1 & 10 & 37 & 77  \\
  \cline{2-7}
        & $n^f$  & 10 & 15 & 56 & 91 &   \\
  \hline
  $d=2$ & $n^E$ &  & 1 & 50 & 83 & 97  \\
  \cline{2-7}
        & $n^f$  & 2 & 54 & 83 & 97 &  
\end{tabular}
\label{tab:torus_convergence}
}
\caption{Convergence of modified Laughlin states with $d=1$ and $d=2$ for $N=10$
  particles. The table contains the number of eigenstates $n^E$ ($n^f$) at
  which the which the energy estimate (cumulative overlap) converges to within 
  $10^{-m}$ of its limiting value (the $n=100$ value in this case). \subref{tab:sphere_convergence} gives values for the sphere and \subref{tab:torus_convergence} for the square torus.}
\label{tab:convergence}
\end{table}
\end{center}

The greatest impact of the fact that $f<1$ is that the overall scale of the overlaps $c_j$ is lower for these states than for the $d=0$ Laughlin state; overlaps for $d=0$ are included in the figures for comparison.
Note that the $c_n$ still fall off rapidly as a function of $n$, so that the
bulk of the $d=1$ state is captured using as few as 20 states on the torus and
fewer still on the sphere. The energies of the $d=1$ states stabilize in a
similar manner to the LLL--content.
For $d=2$ (lower panels) more states are needed before $f$ has converged. 
On the torus, as many as $80\sim100$ states are now needed to reach stable values of $f$ and $E$.
 Nevertheless the number of states needed to capture the $d=2$ state at high accuracy is clearly much smaller than the full Hilbert space dimension of $10^6$ states.
 Similar plots for higher $d$ reveal lower limiting values for $f$ (see \figref{fig:LLL_content_vs_d_N}), but interestingly, the number of states needed for stability of $E$ and $f$ does not increase much beyond what is shown for $d=2$. 

Note in these plots, as is generic for the method, that we are only able to resolve overlaps down to some finite size set by the number of MC samples.
 This scale is set at overlaps of \eg size $10^{-4}$ on the torus for $N=10$ and $2\times 10^7$ MC samples, whereas it is at \eg $10^{-5}$ on the sphere for $N=10$ and $1.2\times10^8$ MC samples.
On the torus, this can be directly observed from the band of low overlaps in the plots.
 These represent the zero overlaps of states with $C_4$ symmetry different from the modified Laughlin wave functions.
These states could be excluded from the analysis, but only for the square torus. In other geometries these states would all have
non-zero overlap and could contain important information on the
reconstruction of the state being projected.

Many other tests of the energy projection could be devised.
 Most obviously one may apply it to other classes of well known wave functions.
 We have done this for example for a number of composite fermion or hierarchy states and the results are qualitatively similar to those for the (modified) Laughlin wave functions.
 One may also calculate the overlap of the unprojected trial wave function with high energy eigenstates of the hamiltonian, to make sure no important components of the LLL-projection at high energy are missed.
 Clearly for large systems this can only be done for a number of eigenstates that is much smaller than Hilbert space dimension, so one would need to have an idea where to look for the potential missing overlap.
 Generally one would observe the behavior for smaller systems to see if there is such high energy overlap and hope that, if there is none, it does not appear in large systems either.  


\begin{figure*}[htb]
\begin{center}
\begin{tabular}{cc}
  Sphere & Torus\\
\includegraphics[width= \picwidth]{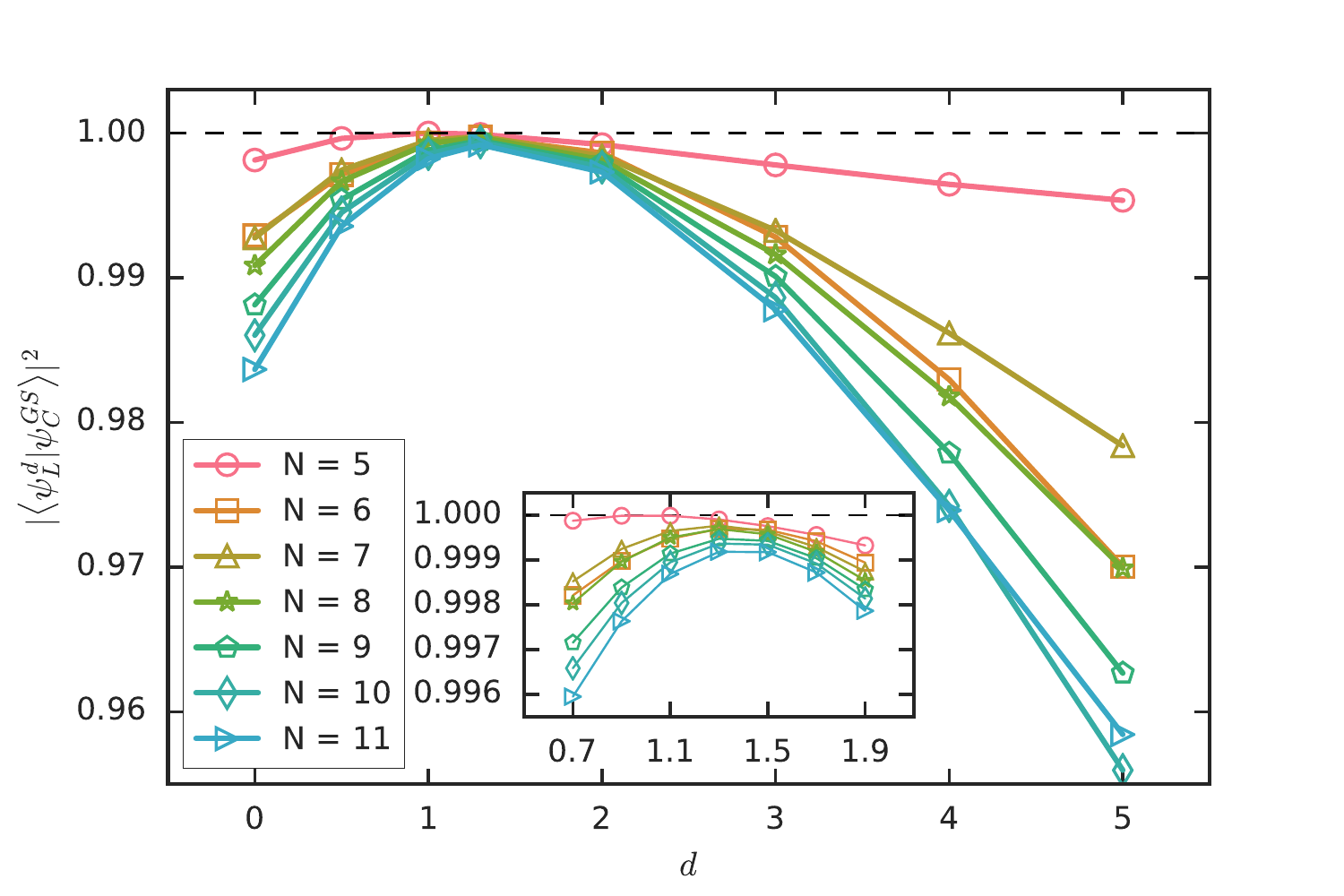}&
\includegraphics[width= \picwidth]{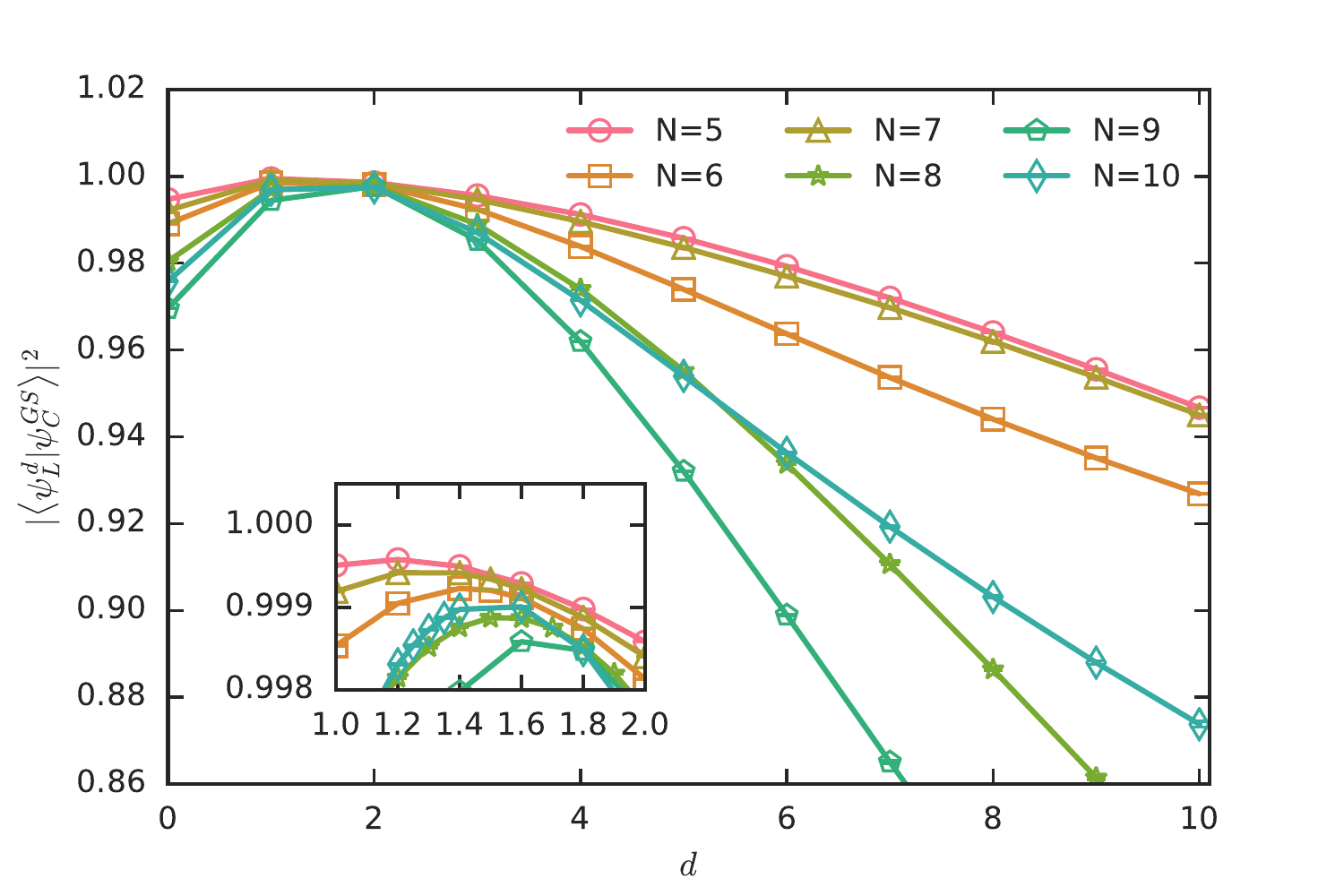}
\end{tabular}
\caption{{\small 
Squared overlap of the modified Laughlin states with the Coulomb ground state,
as a function of $d$ and $N$ on the sphere (left) and torus (right). The insets
zoom in around the optimal values of $d$. Note that errorbars are included but
so small they are not visible. 
}}
\label{fig:ground_state_overlap_scaling}
\end{center}
\end{figure*}

\section{Modified Laughlin states as trial wave functions}

We now turn from testing the energy projection
to using it as a tool to analyze the modified Laughlin states as trial wave functions for the LLL Coulomb problem.
  Since we will be working on the sphere and torus we give explicit expressions for the sphere and torus versions of the states below.
We then go on and study the variational energies and overlaps with the exact Coulomb ground state as a function of $d$ and $N$, as well as two-point correlation functions and entanglement spectra.
 We will find that by letting $1<d<2$, we can significantly improve on the $d=0$ Laughlin wave function.

The explicit form of the modified Laughlin wave functions on the sphere are obtained by directly generalizing the planar wave function from \refcite{Girvin84} to the spherical geometry introduced in \refcite{Haldane83}.
The wave functions on the sphere are 
\begin{equation}
  \psi^{(q,d)}=\prod_{i<j}(u_iv_j-u_jv_i)^q|u_iv_j-u_jv_i|^{2d},    \label{eq:Mod_Laughlin_Sphere}
\end{equation}
written in terms of spinor coordinates $u=\cos(\frac\theta2) \exp(i\frac\phi2)$ and $v=\sin(\frac\theta2) \exp(-i\frac\phi2)$.
Here the spherical coordinates are (radius, polar, azimuthal) $ = (R,\theta,\phi)$, with $R=\sqrt{N_{\Phi}/2}$.

The explicit form of the modified Laughlin wave functions on the torus was introduced in \refcite{Fremling13} and is a natural generalization of the toroidal Laughlin wave function constructed by Haldane and Rezayi in \refcite{Haldane85}.
 The wave functions on the torus (in Landau gauge) are
\begin{eqnarray}
  &&\psi_n^{(q,d)}=e^{-\frac{q+2d}{2q}\sum_{i}y_{i}^{2}}\nonumber\\
  &&\quad\times\prod_{i<j}
  \elliptic 1{z_{ij}}{\tau}^q
  |\elliptic 1{-\bar{z}_{ij}}{-\bar{\tau}}|^{2d}\label{eq:Mod_Laughlin_Torus}\\
  &&\quad\times\ellipticgeneralized{\frac nq+\alpha}{\alpha}{\left(q+d\right)Z-d\bar Z}{\tau\left(q+d\right)-\bar{\tau}d}.\nonumber
\end{eqnarray}
Here $n=1,\ldots,q$ enumerates the different momentum sectors, and $\alpha=\frac 12(N_e-1)$ is chosen for periodic boundary conditions.
We have defined $z_{ij}=\frac{z_i-z_j}{L_x}$ and $Z=\sum_{j=1}^{N_e}\frac{z_i}{L_x}$ to be the relative and center of mass coordinates respectively.
The area of the torus is $L_x L_y=L_x^2\tau_2=2\pi N_\Phi\ell_B$ and the modular parameter $\tau=\tau_1+i\tau_2$ encodes the geometry of the torus.
The torus version of the Jastrow factor consists of $\elliptic1z\tau=\ellipticgeneralized {\frac12}{\frac12}z\tau$ where

\[\ellipticgeneralized a b z\tau = \sum_{k=-\infty}^\infty e^{i\pi\tau(k+a)^2}e^{i2\pi(k+a)(z+b)},\]
  is a generalized Jacobi theta function.
 Since, at small $|z|$, $\elliptic 1 z\tau\approx z\cdot\ellipticprime1 0\tau$, the short distance correlations of \eqref{eq:Mod_Laughlin_Torus} are the same as those of the planar version in \eqref{eq:CF_basic}.

\subsubsection{Coulomb overlap}
We start by considering the overlap with the Coulomb ground state as a function of $d$ and $N$.
 In \figref{fig:ground_state_overlap_scaling} this is shown for both the sphere (left) and torus (right).
The main feature of interest is that the overlap of the standard Laughlin state is systematically improved for all system sizes by tuning $d>0$.
 Values of $d$ between $d=1$ and $d=2$ give the best overlap with the Coulomb ground state.
 While the optimal value of $d$ is not completely independent of system size,
 this dependence is weak (especially on the sphere) and we note that near the optimal value of $d$ the overlap decreases only very slowly with increasing system size.
 For values of $d$ with lower overlaps (and notably for the standard Laughlin wave function at $d=0$), the overlap also decreases much faster with system size.
 The optimal squared overlap is above $0.998$ on the torus for all system sizes considered (up to $N=10$).
 On the sphere the system sizes go even to $N=11$ and we still obtain optimal squared overlap of $0.999$.

Note that these figures show results for many fractional values of $d$.
That we are able to project wave functions that have a fractional value of $d$ is a powerful feature of the energy projection method.
In many other methods this kind of projection would be difficult as it would be unclear how to handle fractional powers on the Jastrow factors.
Here the projection is no more difficult than that of integer $d$, and the only extra effort lies in generating the unprojected wave functions.


\subsubsection{Variational energy}

\begin{figure*}[htb]
\begin{center}
\begin{tabular}{cc}
  Sphere & Torus\\
\includegraphics[width= \picwidth]{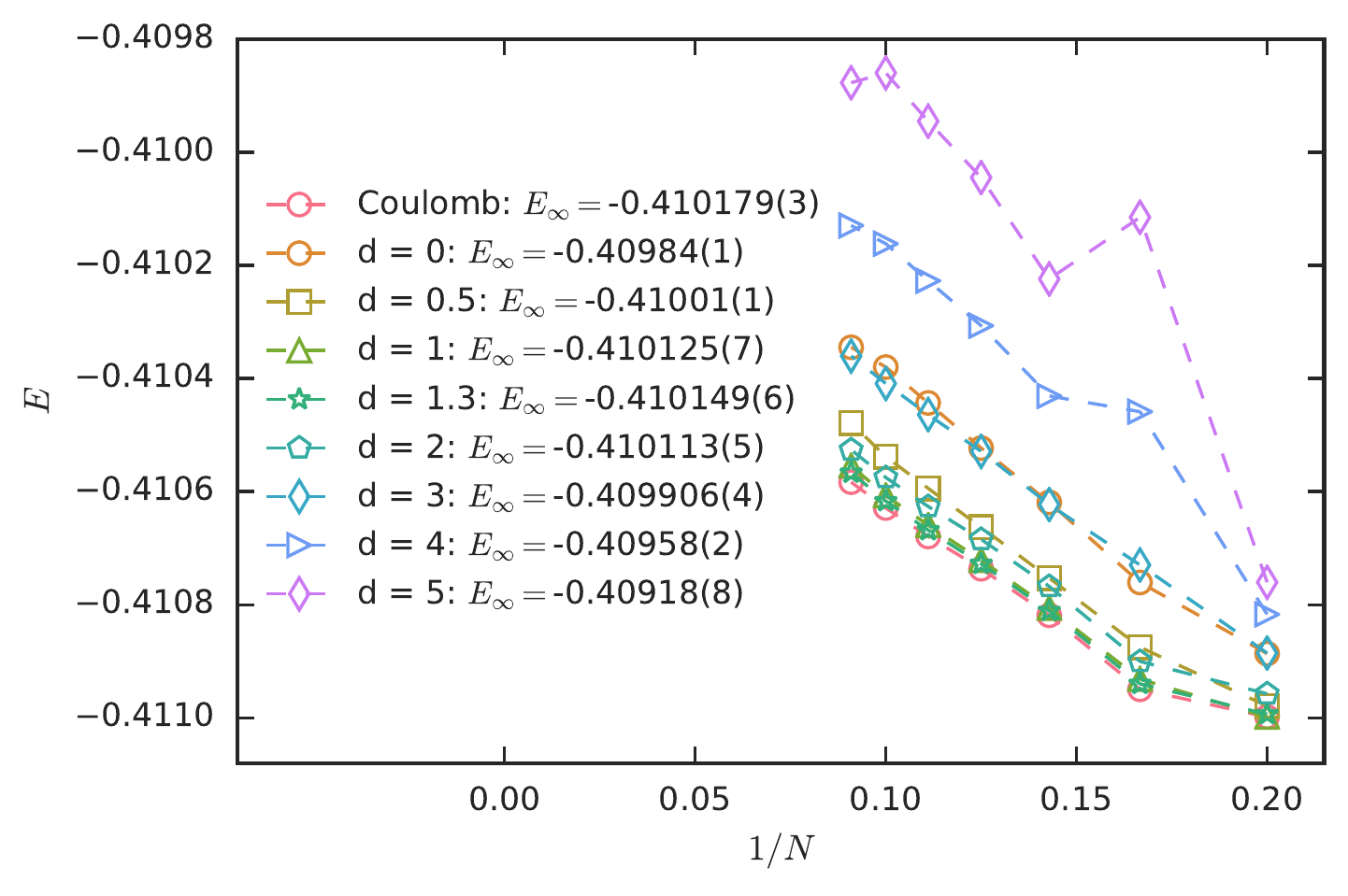}
&
\includegraphics[width= \picwidth]{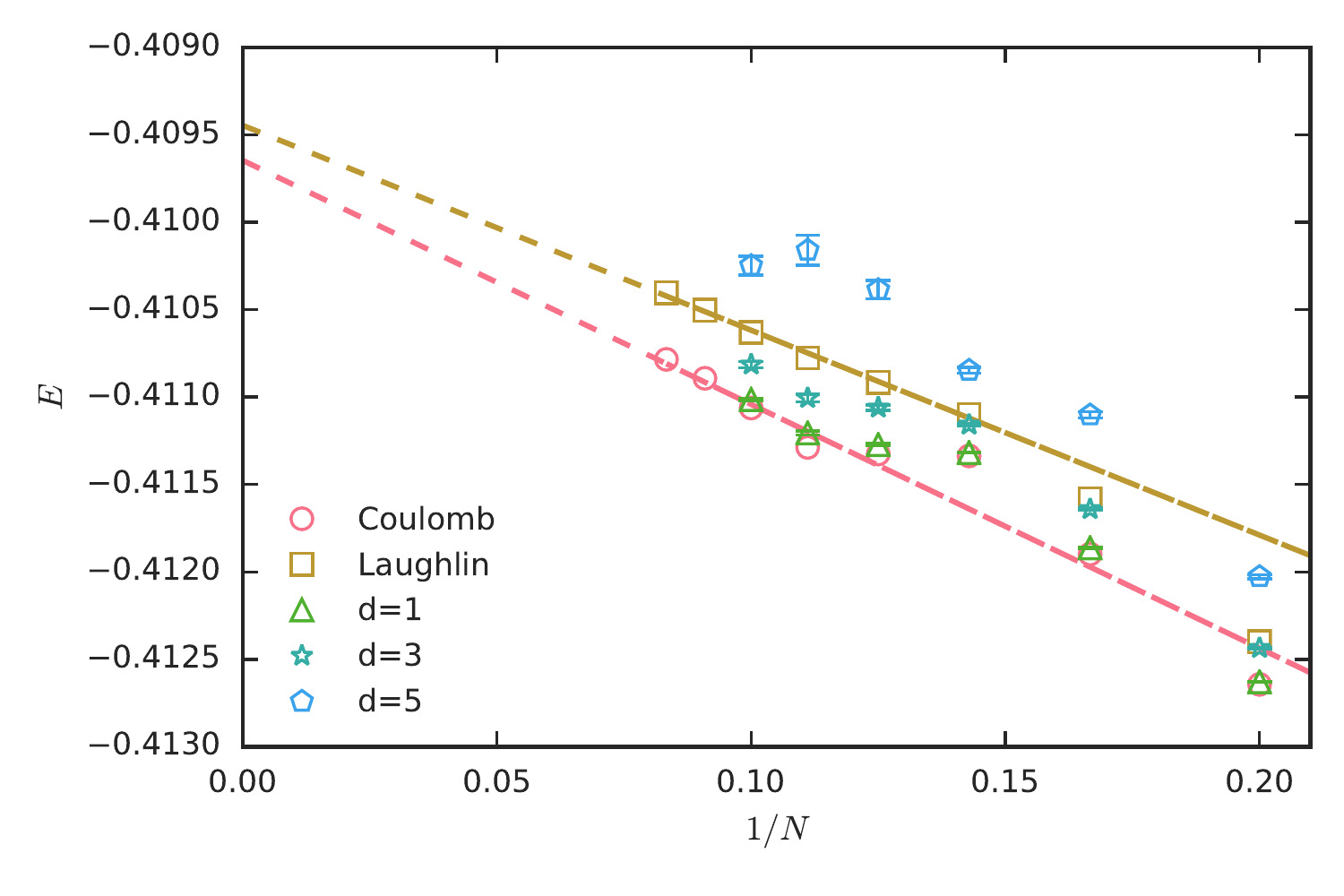}
\end{tabular}
\vspace*{-5mm}
\caption{{\small 
Variational energy per particle with varying system size on the sphere (left) and torus (right), after background subtraction (see main text) for various values of $d$ (indicated in the figure).
Energies plotted against $1/N$ to indicate scaling behavior.
Errors are indicated where they are larger than the symbols used.
The sphere plot shows limit values of the energy as $N\rightarrow\infty$ obtained from linear extrapolations in $1/N$.
The torus plot shows linear extrapolations for the Coulomb and Laughlin states. \\
}}
\label{fig:energy_scaling}
\end{center}
\end{figure*}

Another measure of the quality of a trial ground state is its variational energy.
Results for the variational energy of the modified Laughlin states for various values of $d$ and $N$ are shown in \figref{fig:energy_scaling}.
The energy per particle is plotted against $1/N$ to detect scaling behavior for $N\rightarrow \infty$.
To obtain the correct scaling we perform the usual background subtractions and density corrections on the sphere (see \eg \refcite{Jain_CF}, appendix I) and background subtractions on the torus\cite{Yoshioka83,Bonsall59}.

On the sphere we find energies lower than that of the Laughlin state in the region $0<d<3$ (for all $N$).
The minimal energies at these finite system sizes are found around $d=1.3$.
The energies for $0<d<3$ appear to be in a scaling region for systems from size $N=7$ upwards, enabling an attempt at computing the thermodynamic ground state energy density.
For the modified Laughlin state at $d=1.3$ we thus obtain a variational energy per particle $E_{d=1.3}=-0.410149(6)$.
This should be compared to the scaled Coulomb energy at $E_C=-0.410179(3)$ and the scaled Laughlin energy at $E_L=-0.40984(1)$.
Limiting values of the variational energy for other values of $d$ are given in the figure. 
The reported errors give one standard deviation, and only reflect the uncertainty that comes from MC estimation and the linear fit.
This leaves out effects from the cutoff in energy eigenstates used in the projection and more importantly any finite size effects which may still occur at larger system sizes.
Nevertheless, it is clear that the modified Laughlin state at $d=1.3$ has an excess energy which is an order of magnitude smaller than that of the standard Laughlin state at all finite sizes considered and we expect this will continue to be true at larger sizes.

On the torus, finite size effects are larger and we have not performed fits of the energy for $N\rightarrow \infty$, except for the Coulomb and standard Laughlin states, which we calculated out to larger sizes than the energy projected modified Laughlin states.
Because the points jump around more we have only included $d=1$, $d=3$ and $d=5$ results in this plot, to keep it readable.
Nevertheless, the general picture is more or less the same as on the sphere, in that modified Laughlin states with $1<d<2$ can very significantly reduce the variational energy from that of the standard Laughlin wave function at all system sizes examined.
For example at $N=10$, the energy obtained for $d=1.3$ is $E_{d=1.3}=-0.41061$ as compared to the Coulomb energy at $E_C=-0.41063$ and the Laughlin energy of $E_L=-0.4104$.

\begin{figure*}[htb]
\begin{center}
\begin{tabular}{cc}
  Sphere & Torus\\
\includegraphics[width= \picwidth]{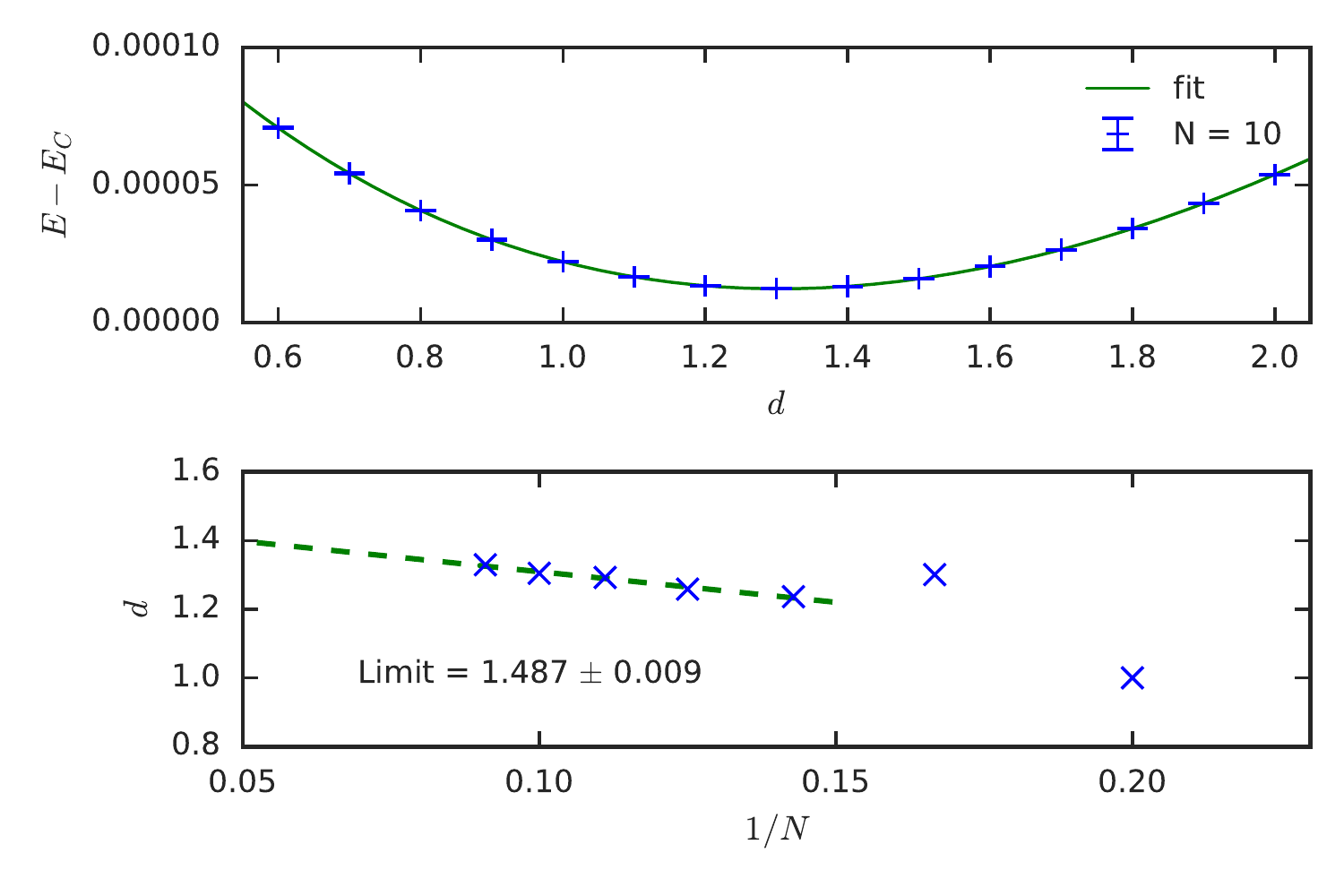}&
\includegraphics[width= \picwidth]{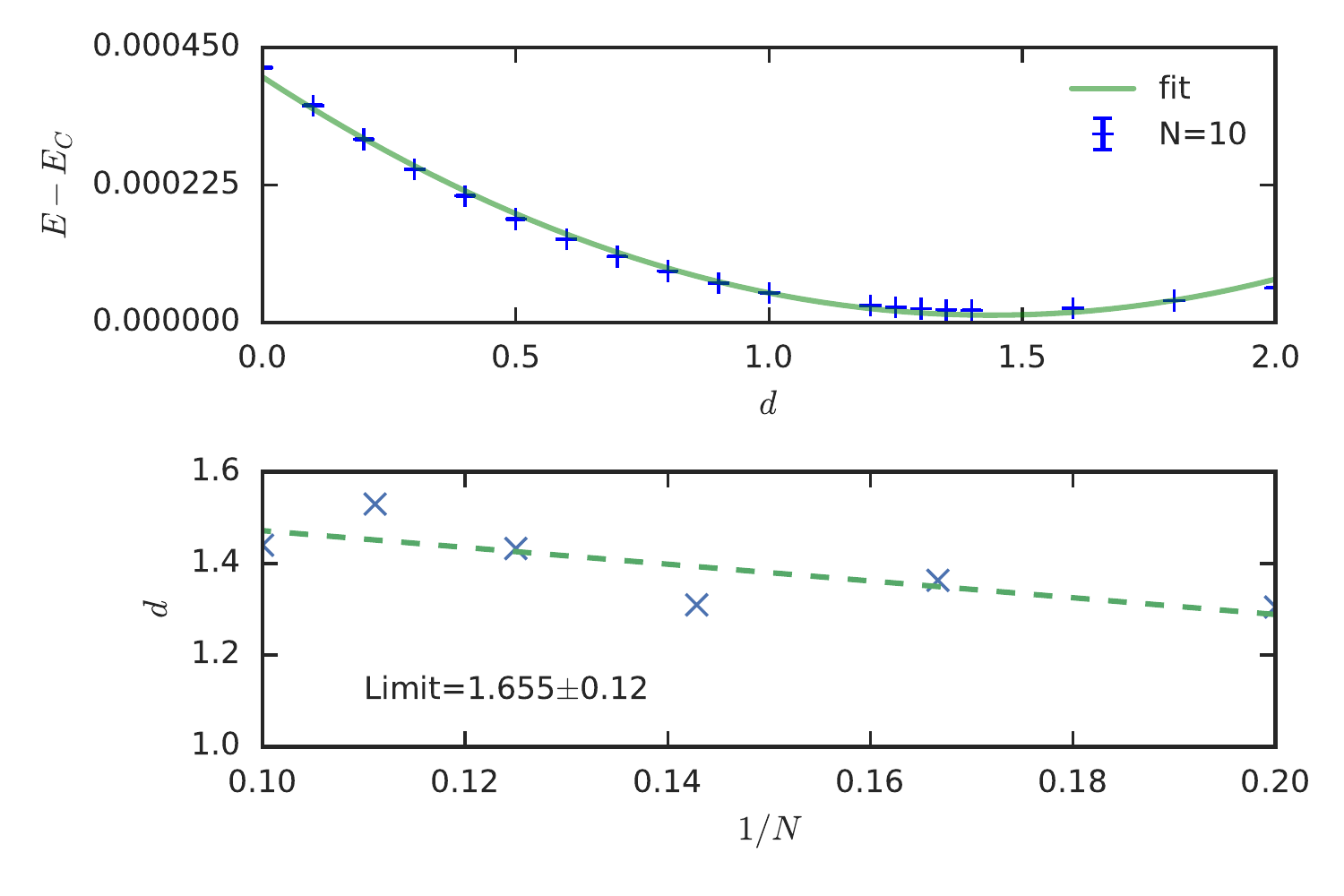}
\end{tabular}
\caption{{\small 
\textbf{Upper panels:}
Variational LLL Coulomb energy per particle of the modified Laughlin states as a function of $d$ for $N=10$ particles on a sphere (left) and on a torus (right).
The exact Coulomb energy is subtracted.
The plots show clearly that a very significant reduction of the variational energy is obtained upon changing from the Laughlin (d=0) state to the modification with the optimal value of $d$.\\
\textbf{Lower panels:}
Scaling with $N$ of the value of $d$ where the minimum energy is obtained.
The minimum $d$-values at various $N$ are obtained from parabolic fits such as those shown in the upper panels.
}}
\label{fig:GS_energy_scan_of_d}
\end{center}
\end{figure*}

In \figref{fig:GS_energy_scan_of_d}, we show the variational energy on the sphere (left) and torus (right) as a function of $d$, for $N=10$ (upper panels).
It is clear that the energy is a smooth function of $d$ which is very well fit by a low order polynomial.
We use the value of $d$ where this fit takes its minimum as a good estimate for the optimal $d$ at a given $N$.
These optimal values of $d$ were plotted against $1/N$ (lower panels) to get an idea of the best possible value of $d$ in the thermodynamic limit.
On the sphere there is again what appears to be excellent scaling behavior from $N=7$ upwards, leading to an estimated limit value $d_{\infty}=1.487\pm 0.009$. On the torus, finite size effects again appear larger, but a linear scaling fit can still be attempted leading in this case to $d_{\infty}=1.655\pm 0.12$.
Again, the errors on these numbers represent a single standard deviation and do not take into account finite size effects which may manifest when considering larger sizes.
It is encouraging that there appears to be proper scaling behavior of $d$, as this supports the idea that $d$ is a physical parameter of the system in the thermodynamic limit. 

\begin{figure*}[htb]
\begin{center}
\begin{tabular}{cc}
  Sphere & Torus\\
  \includegraphics[width= \picwidth,clip=true,trim={0 0 2cm 0}]{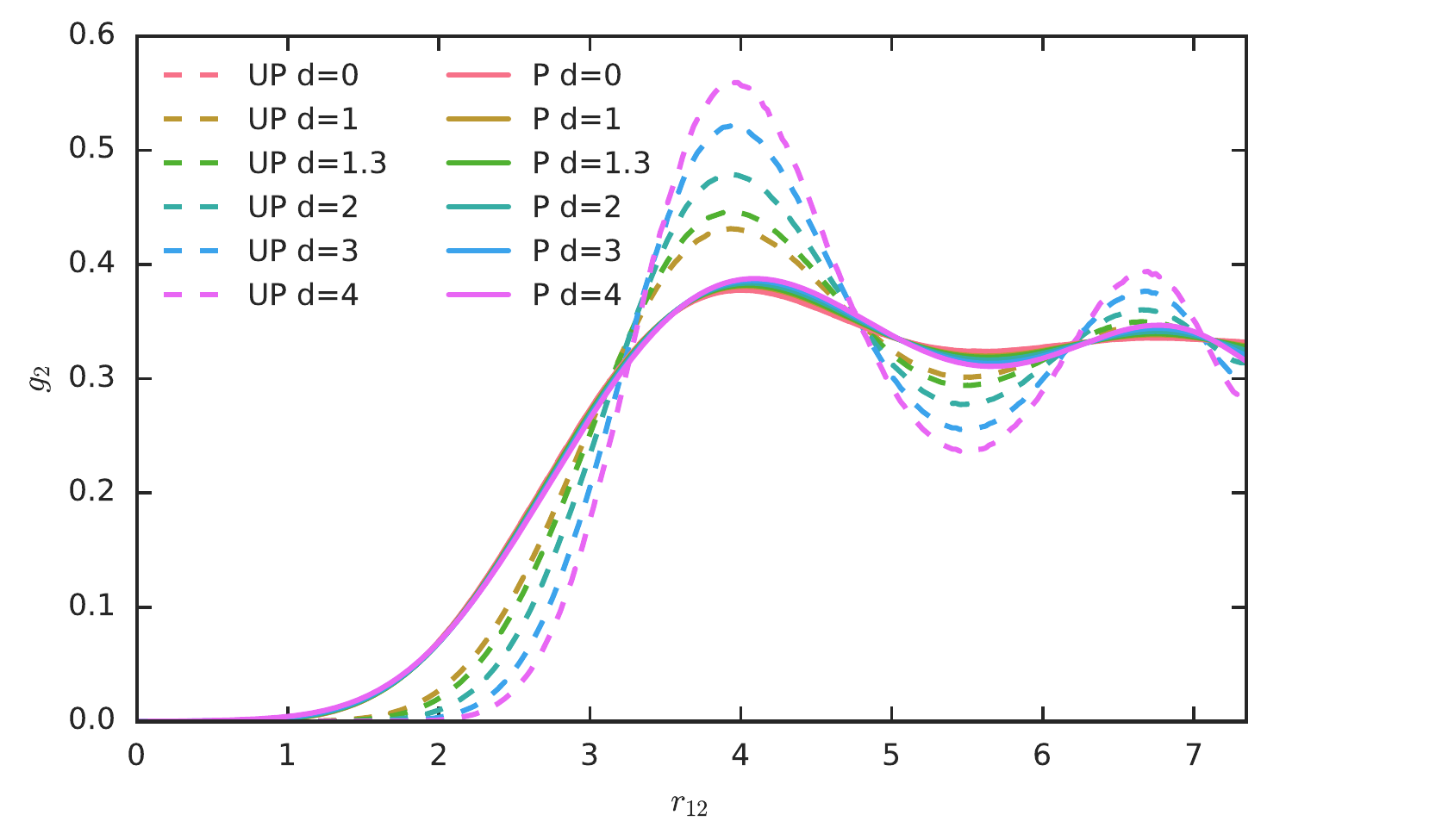}&
  \includegraphics[width= \picwidth]{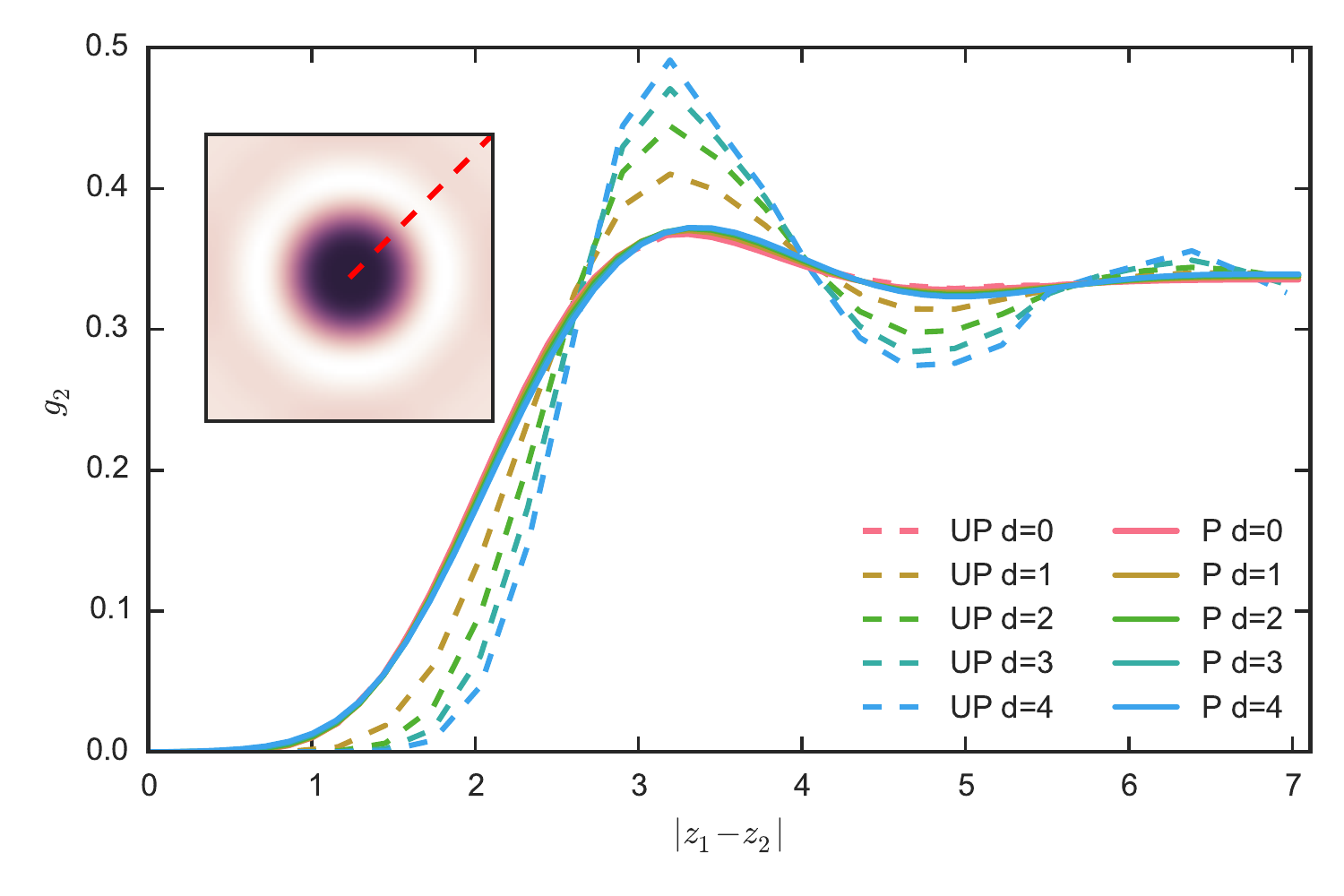}\\
  
  \includegraphics[width= \picwidth]{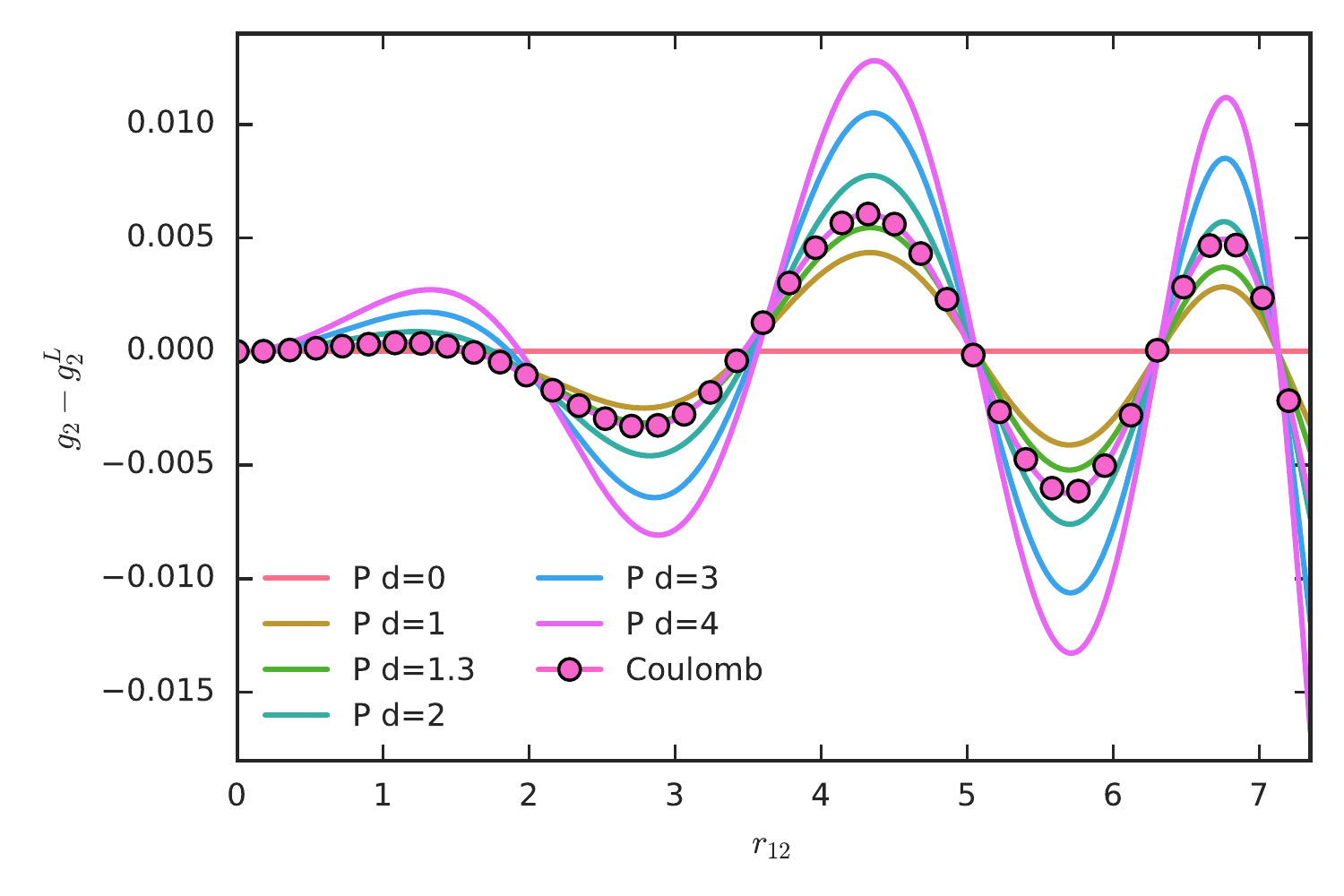}&
  \includegraphics[width= \picwidth]{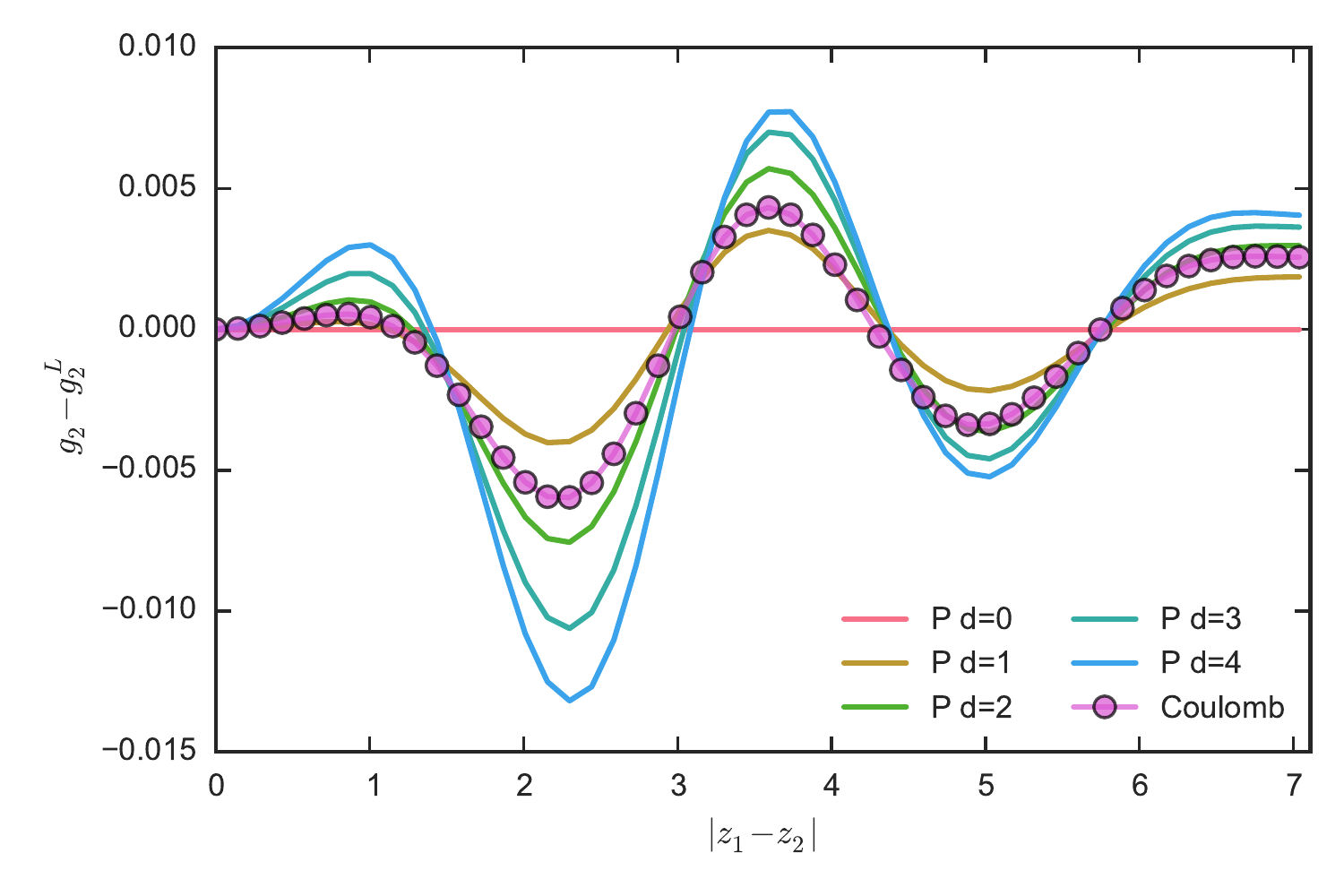}
\end{tabular}
\caption{\small{Two-particle correlation function of the modified Laughlin wave functions on the sphere (left panels) and torus (right panels) for $N=10$ particles.\\
\textbf{Upper panels:} Dashed lines show the correlation function of the unprojected alternative Laughlin wave functions. 
Solid lines show the correlations after projection to the LLL.
On the sphere, the distance measure is the chord length $r_{12}=2R|u_1v_2-u_2v_1|.$\\
On the torus, the curve represents a diagonal cut of the square torus.
The inset is a plot for the full torus, indicating the cut shown in the main panel.\\
\textbf{Lower panels:} The same functions as in the upper panels, but with the correlation function for the ($d=0$) Laughlin wave function subtracted,
to highlight the differences between Coulomb, Laughlin and modified Laughlin correlations.}}
\label{fig:correlation}
\end{center}
\end{figure*}

\subsubsection{Two-particle correlation functions}
The intuition which led Girvin and Jach\cite{Girvin84} to introduce the modified Laughlin wave functions was that a nonzero $d$ would ``discourage close encounters of the particles".
While this seems obvious for the unprojected wave function, it is less obvious after projection.
For example we may observe that the planar wave functions (\ref{eq:alt-L_basic}) are all the same for $N=2$ (after projection).
Also, our results on the variational energy show that while close encounters may be discouraged for $1<d<3$, this is not so clear for large $d$, where the variational energy increases again.
To directly investigate the matter we have calculated the $2$-particle correlation functions of the modified Laughlin states. 

Correlation functions are shown in \figref{fig:correlation} both for the sphere (left panels)  and for the torus (right panels). 
The correlation  function on the sphere depends only on the distance between the particles.
We can think of it as a density plot for a system with one particle fixed at the north pole.
The plot for the (square) torus has the first particle fixed at $z=0$ and is showing a diagonal cut to $z=\frac{1+i}2L$ (the diametrically opposed point of the square),
with a density plot of the full 2D correlation function in the inset.
In the upper panels of the figures, the correlations for the unprojected wave functions are shown as dashed lines for $d\in\{1,2,3,4\}$.
In these plots we see two very clear trends with increasing $d$:
the correlation hole around $z=0$ widens, showing directly that close encounters are discouraged before projection,
and the oscillations at larger distances increase, showing increasing signs of the local onset of crystallization. 
We note that the wave length of the oscillations appears to be fairly independent of $d$ and is approximately $1.5\ell_B$.
The correlation functions after projection are shown as solid lines in the same figures.
We see that almost the entire effect observed before projection is reversed.
This is likely due to the fact that the basis functions of the LLL only allow particles to be localized to within about a magnetic length,
which limits the sharpness of any peaks in the correlation function.
In order to see the remaining modifications clearly,
we plot the difference between the correlation functions for the same $d$'s and the standard Laughlin wave function in the lower panels,
as well as the difference between the correlation function for the Coulomb ground state and the Laughlin state.
We see that the remaining effects still echo the effects observed before projection.
As $d$ increases the correlation functions have increasing oscillatory behavior at the same wave length as before projection.
The widening of the correlation hole is now seen to be simply part of this oscillatory behavior.
On the sphere we note that the strongest effects of the modification are at longer distances with the correlation hole less affected.
Note that the use of chord length in \figref{fig:correlation}, instead of arc length,
makes the oscillations at larger $r_{12}$ on the sphere appear to have shorter wavelength than is actually the case.
We also see, in good agreement with what we know from the overlap and energy,
that the best fit to the Coulomb correlation function lies somewhere in between $d=1$ and $d=2$.
The Coulomb ground state clearly has stronger long range oscillations than $\psi_L$,
which fits with intuition, since $\psi_{L}$ is the ground state of an ultra short ranged interaction,
while the Coulomb interaction is long ranged.
We can think of the introduction of a nonzero $d$ as a way to reintroduce these longer range oscillations. 

For the sphere we have also added the curve for $g_{d=1.3}$.
We see that $g_{d=1.3}\approx g_C$ to very good accuracy,
especially at shorter distances, where the difference is imperceptible in the plot. 
At the longest distances, $g_C$ is closer to $g_{d=2}$.
Perhaps this is related to the fact that $d$ drifts towards $1.5$ at large $N$, where longer distances can be probed.
Finally, we note that while close encounters may be discouraged for $1<d<3$, For large d, the correlation functions start to show an increased probability to find pairs of particles at a distance of approximately $1$ magnetic length (where each particle would
be inside the other's ``correlation hole'').
\begin{figure*}[htb]
  \begin{center}
    Entanglement spectra sphere
    \begin{tabular}{cc}
      \includegraphics[width= \picwidth]{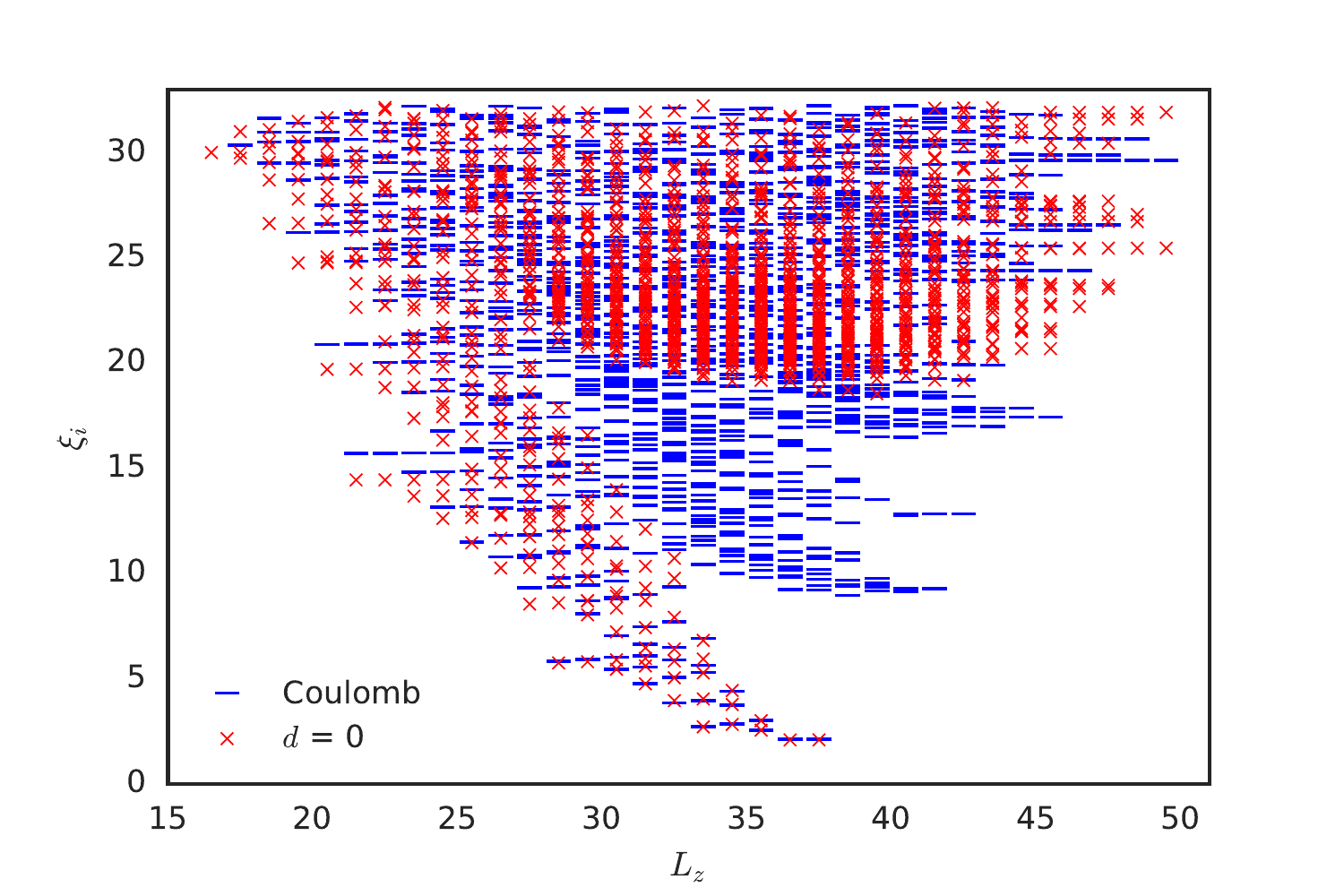}&
      \includegraphics[width= \picwidth]{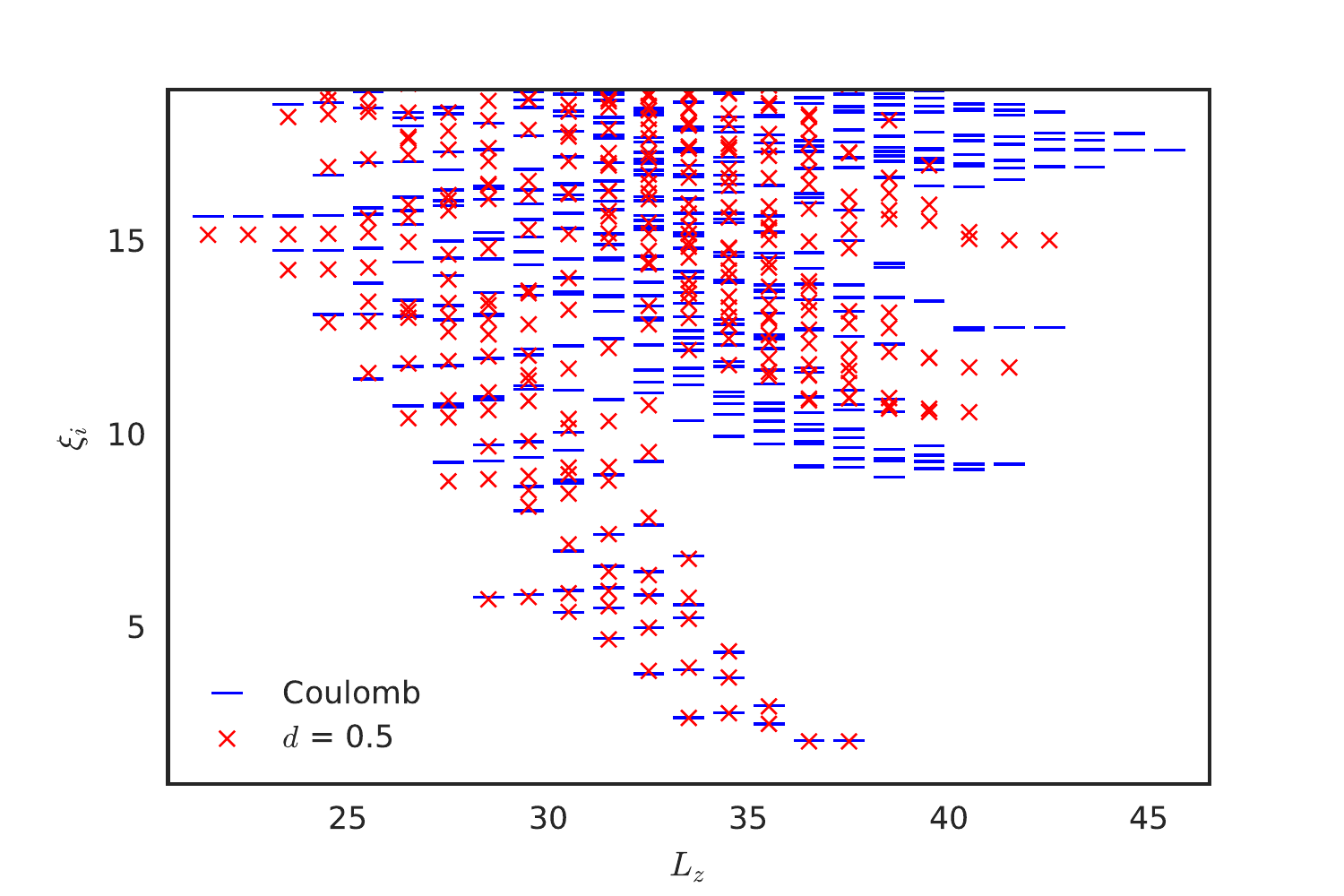}\\
      \includegraphics[width= \picwidth]{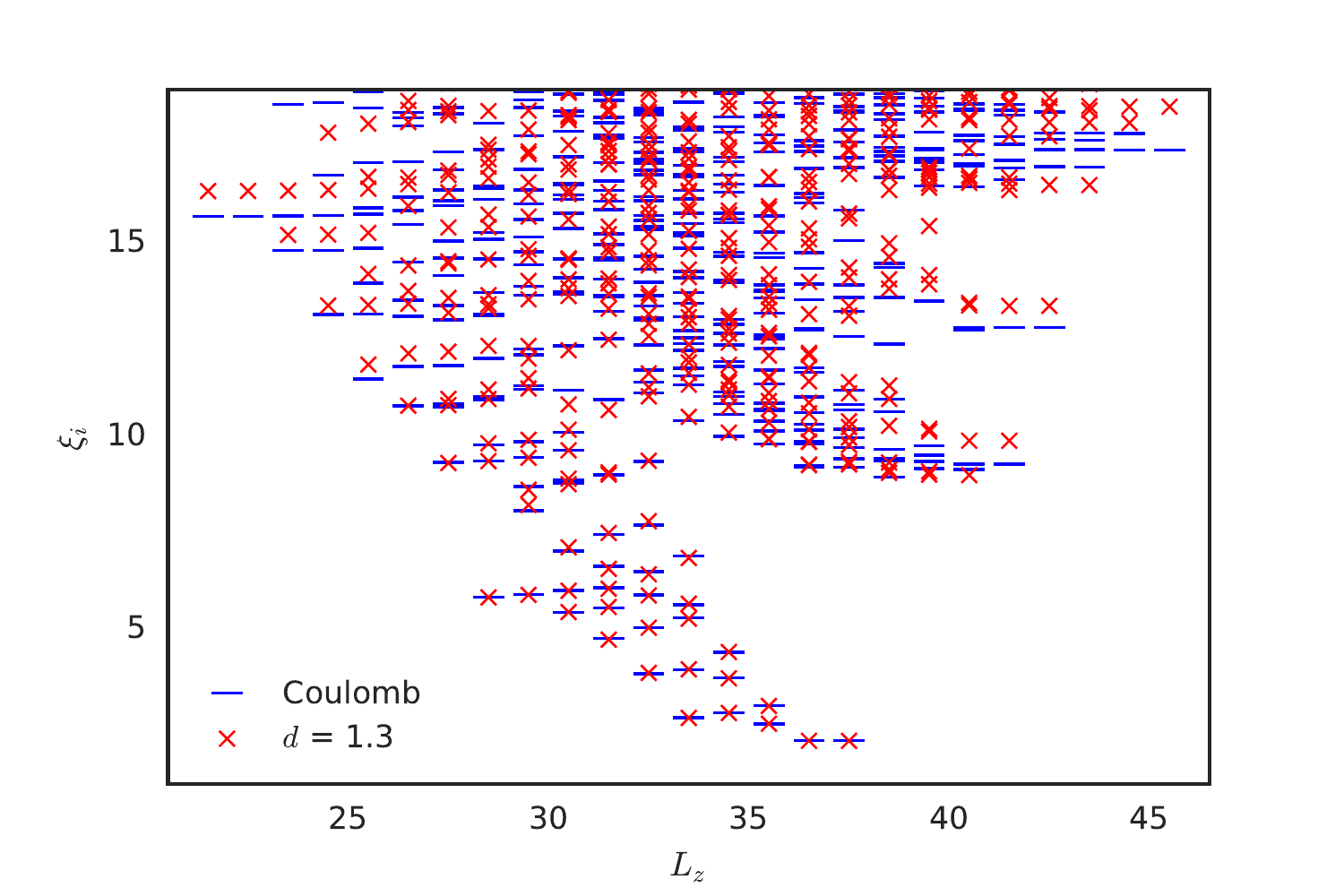}&
      \includegraphics[width= \picwidth]{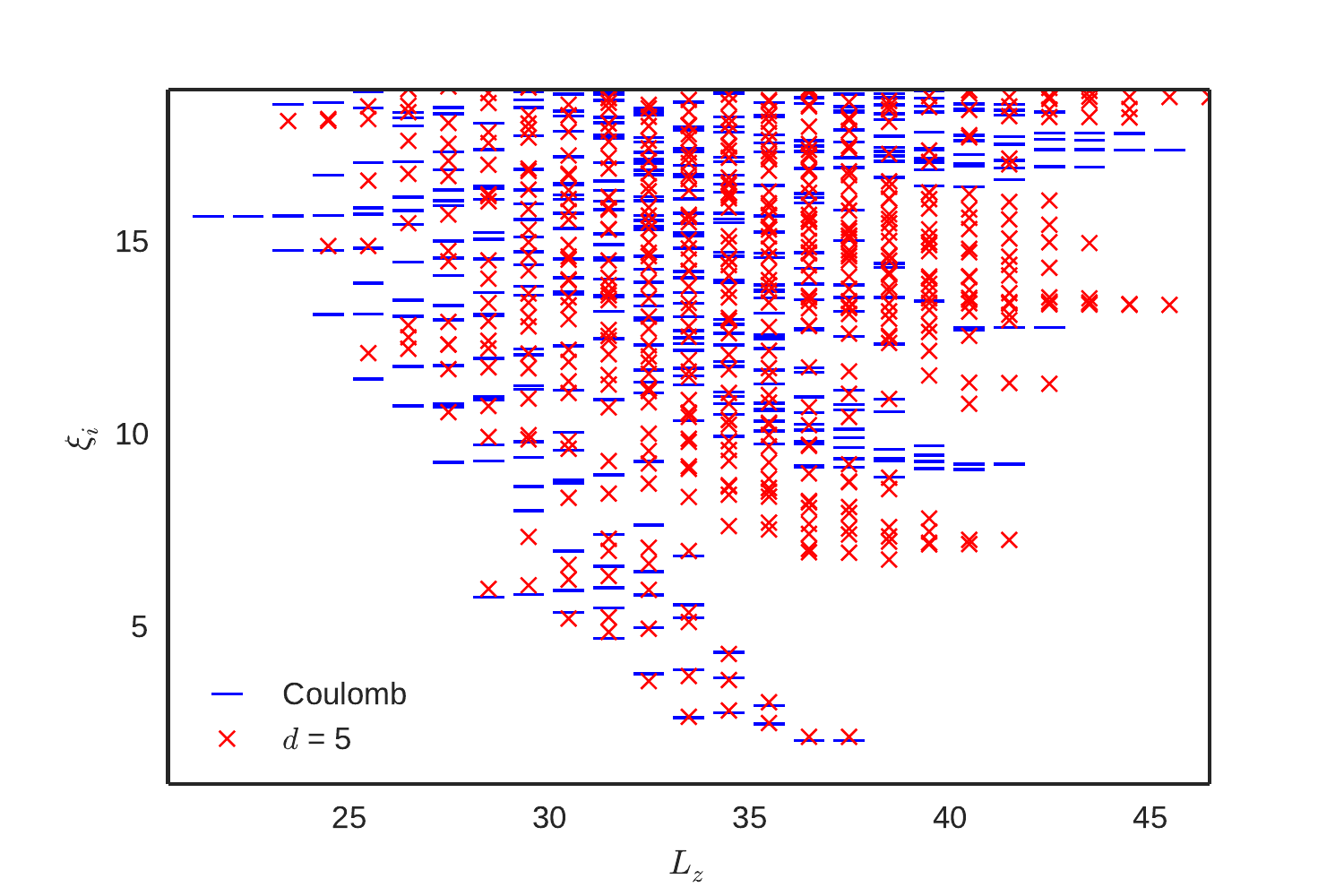}
    \end{tabular}
    \caption{{\small 
        Entanglement spectra for $N=10$ particles on a sphere as function of $d=0$ (upper left), $d=1$ (upper right), $d=1.3$ (lower left) and $d=5$ (lower right).  All figures also show the Coulomb entanglement spectrum (dashes) for comparison.
    }}
    \label{fig:ES_sphere}
  \end{center}
\end{figure*}

\begin{figure*}[htb]
  \begin{center}
    Entanglement spectra torus
    \begin{tabular}{ccc}
      \includegraphics[width=6cm]{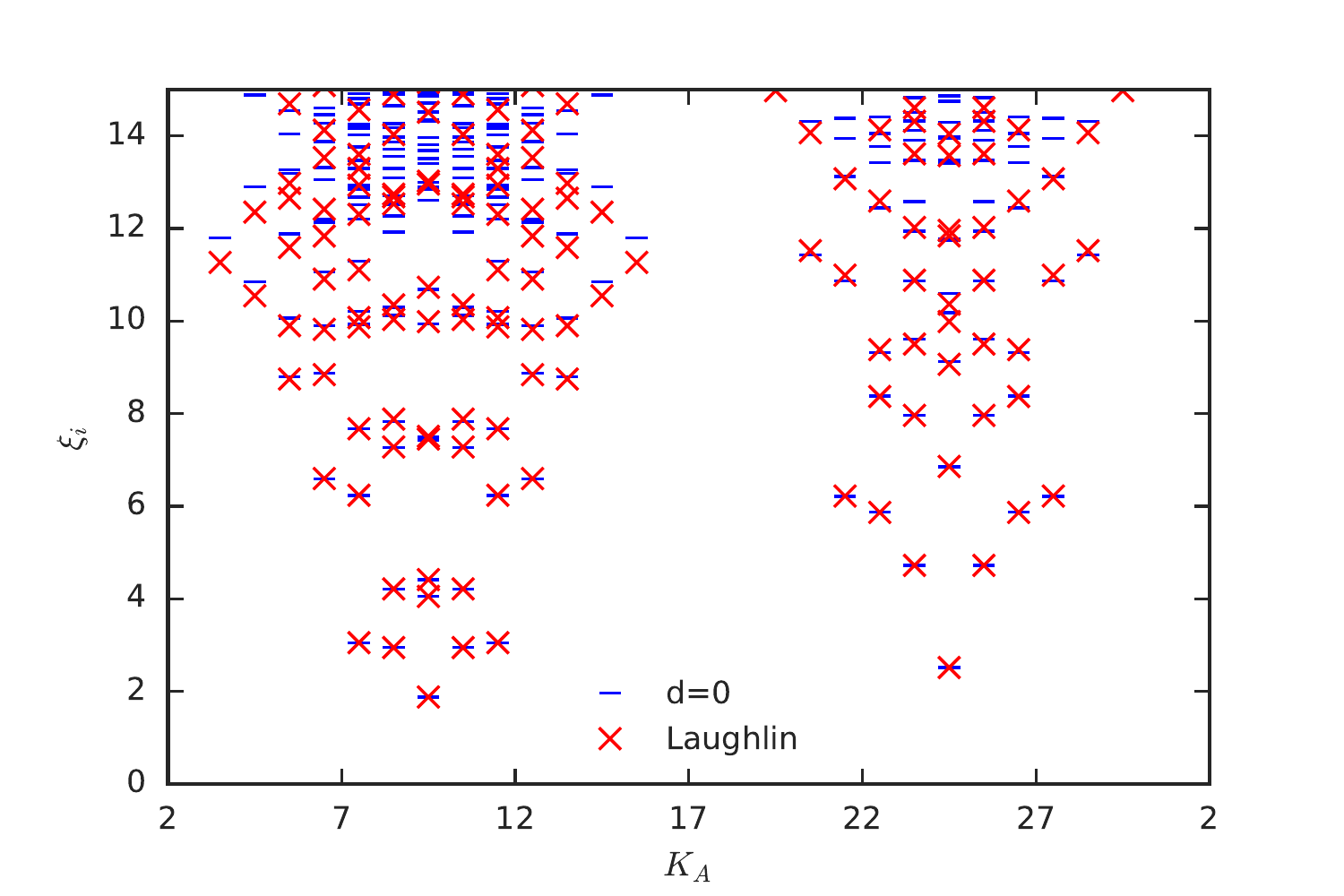}&
      \includegraphics[width=6cm]{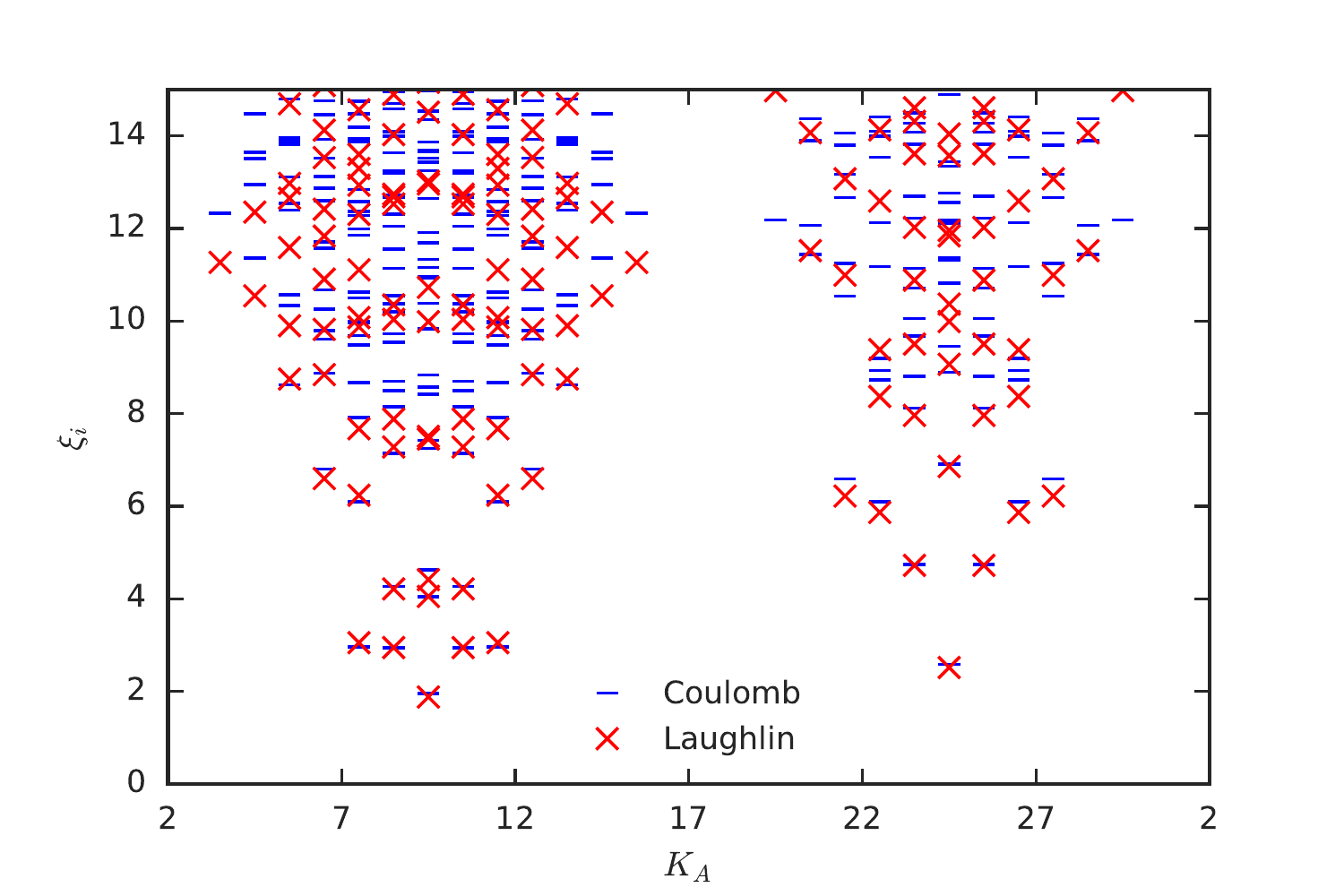}&
      \includegraphics[width=6cm]{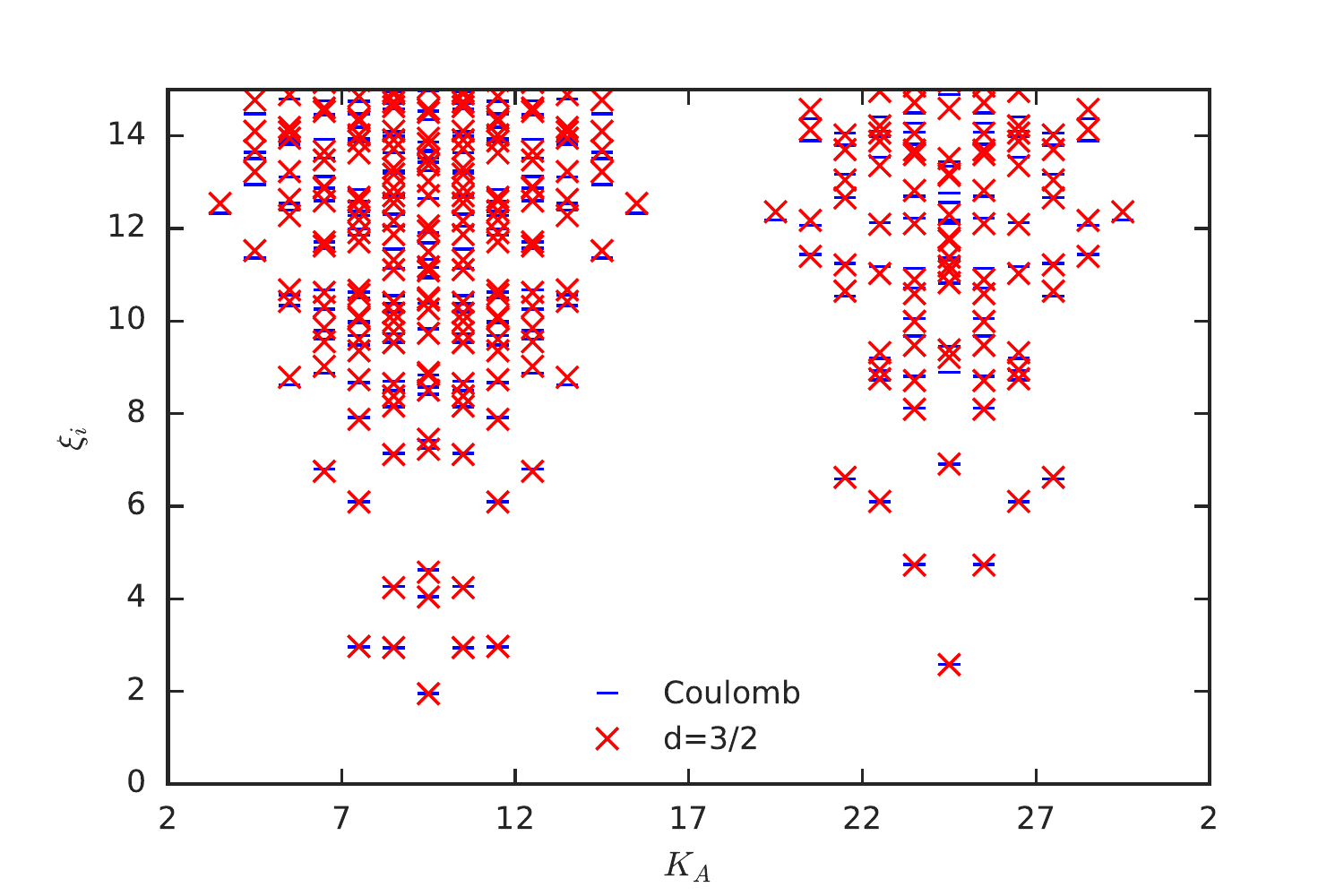}
    \end{tabular}
    \caption{{\small 
        Entanglement spectra for $N=10$ particles on a torus\\
        \textbf{Left panel:} ES of the exact Laughlin state (crosses) superimposed on the ES for the energy projected $d=0$ state (dashes).
        The difference between these spectra is due to Monte Carlo error in determining the $d=0$ state.
        We see that the projected data cannot be trusted for entanglement energies above $\xi=12$.\\
        \textbf{Middle panel:} Comparison of exact Coulomb (dashes) and Laughlin spectra (crosses).
        We see that, as in the case of the sphere ES, many states are missing from the Laughlin ES at entanglement energies above $\xi\approx 8$.\\
        \textbf{Right panel:}  ES of the $d=1.5$ modified Laughlin state (crosses) superimposed on the exact Coulomb ES (dashes).
        The missing states in the Laughlin ES are accounted for, and the fit on the other states are also improved over the middle panel. 
    }}
    \label{fig:ES_torus}
  \end{center}
\end{figure*}

\subsubsection{Entanglement spectra}

Finally, we consider the entanglement spectrum\cite{Li08} (ES) of the modified Laughlin states, using the orbital cut introduced in \refcite{Haque07}.
The entanglement spectrum is a powerful tool for the determination of the topological order of gapped systems.
For Hall states, the low lying part of the ES of a system on the sphere often resembles the spectrum of the chiral conformal field theory (CFT) describing the modes propagating along the circular edge of the corresponding state on a disk\cite{Li08}. 
For Hall systems on the torus, the ES will resemble the edge spectrum on a cylinder, where the edge consists of two circles governed by counterpropagating versions of the same chiral CFT\cite{Lauchli10}.
The Laughlin state has the special property that \emph{all} states in its orbital ES correspond to states in its edge CFT.
The ES of the exact Coulomb ground state on the other hand has a clearly identifiable low lying branch corresponding to the ES of the Laughlin state, but in addition has many other states in higher branches.
These states can be attributed to components of the Coulomb ground state which can be thought of as neutral bulk excitations of the Laughlin state\cite{Sterdyniak11_NewJPhys}. 

In \figref{fig:ES_sphere} we show the orbital ES of a number of systems of $N=10$ particles on a sphere.
In all four panels, we show the ES of the Coulomb ground state (blue dashes) with superimposed on it the ES of the energy projected state (red crosses).
The top left figure shows the ES of the Laughlin state as determined from its energy projection (we can think of it as the $d=0$ state).
We clearly see from the graph that the Laughlin state indeed reproduces the lowest branch of the Coulomb ES but not the higher ones.
We also see that at values of the entanglement energy $\xi$ above $20$ there are many spurious states in the $d=0$ ES which would not appear if we had used the exact Laughlin state rather than its energy projection.
These states appear purely due to the error of the energy projection.
The scale at which they first appear can in principle be shifted upward by taking more MC samples.
It is clear that with the amount of MC samples we have taken here, states with $\xi>20$ can be safely discarded as noise and we have therefore cut off the scale at this level in the other panels of \figref{fig:ES_sphere}.
In the upper right panel, we consider the ES of the $d=0.5$ modified Laughlin state.
We see that there is still a good fit to the $d=0$ branch of the ES but additional branches of states are swooping down from above as a result of setting $d>0$.
In the bottom left panel we consider $d=1.3$ which gives more or less the optimal fit to the Coulomb energy as well as the highest overlap at this system size.
We see that the branches of the ES have now settled very closely to the location where they are in the Coulomb ES.
The fit of the low lying ES (in the Laughlin branch) is also noticeably improved for $d=1.3$
for entanglement energies up to $\xi\approx10$.
 While the detailed positioning of the individual levels within the higher branches does not always match very well, we stress that the structure of these branches,
\ie the counting of levels at each angular momentum, is identical to that of the Coulomb ES, even though this may not always be obvious from the plot.
We also notice that while the overall trend in raising $d$ has been to bring levels down out of the noise, there are some exceptions at low angular momentum, where the entanglement energies of some levels have risen from values below to values above those of the Coulomb ES.
These trends continue for higher values of $d$.
The ES for $d=5$ is shown in the bottom right panel.
Even at $d=5$ the lower part of the Laughlin branch of the ES is still mostly in place, except for some levels at low angular momentum which seem to have migrated up into the noise.
On the other hand the higher branches visible at large angular momenta have now all descended well below the corresponding Coulomb branches.   

Entanglement spectra for $N=10$ electrons on the torus are shown \figref{fig:ES_torus}.
In the left panel we again compare the energy projected $d=0$ state to the exact Laughlin state to give an idea of the Monte Carlo noise on the data for the modified Laughlin wave functions. Any disagreement between the $d=0$ and exact Laughlin states' ES is due to the error in the determination of the $d=0$ state, which would be exactly equal to the Laughlin state if this error was zero. We see that the noise in the ES in this case becomes severe above entanglement energy $\xi\approx 12$. This more severe noise, as compared to the sphere ES, is due to the fact that the data is based on fewer MC samples. 
In the middle panel, we show the exact Laughlin ES superimposed on the ES for the exact Coulomb ground state.
As on the sphere, there is good matching of the low lying levels, but many
higher lying levels from $\xi\approx 8$ upwards are completely missing from the
Laughlin ES. In the right hand panel, we show the ES for $d=1.5$, which is close
to optimal.  We find that all levels in the Coulomb ES are now reproduced with
excellent matching of the entanglement energies. The entanglement energies
obtained for levels that were already present in the Coulomb ES is also visibly
improved.  Overall, the ES for the torus shows similar feature to those for the sphere.
As $d$ is increased, levels which come from the higher branches of the Coulomb ES come down.
There is also a tendency for levels that are far from the center of the conformal towers of states to be shifted up, which is analogous to the shifting up of high angular momentum states on the sphere.

\section{Discussion}
\label{sec:discussion}

We have introduced the energy projection (EP) as a method for projecting quantum Hall trial wave functions to the lowest Landau level.
In effect we replace the LLL projection by projection onto the low lying spectrum of a suitable hamiltonian acting in the LLL, carefully checking convergence of this projection to what should be the full LLL projection.
The method works well for all states we have considered, up to system sizes where a few hundred states are accessible by exact diagonalization.
We have shown this here in some detail for the Laughlin state and for the modified Laughlin states proposed by Girvin and Jach in \refcite{Girvin84}. 

We have also applied EP to investigate the modified Laughlin states as trial wave functions for the Coulomb ground state at filling $\nu=\frac{1}{3}$.
It turns out that these states allow for significant improvements over the standard Laughlin state.
For example the squared overlap with the Coulomb ground state of a system of $N=11$ electrons on the sphere is improved from $\sim0.98$ for the Laughlin state to $\sim0.999$ for the modified state at $d=1.3$.
On the torus at $N=10$, there is a similar improvement from $\sim0.97$ for the Laughlin state to values above $0.998$ for the modified states with $1.4<d<1.9$.
The variational energy per particle can also be improved from that of the Laughlin state at the finite sizes we considered and likely also in the thermodynamic limit as can be seen from the scaling results in \figref{fig:energy_scaling}.
We also investigated the two particle correlation functions of the modified states and found that, compared to the standard Laughlin state, the states at $d>0$ exhibit more pronounced medium range oscillations, which allows them to better mimic the Coulomb ground state.
While close encounters of the particles are to some extent discouraged at $d>0$
(as expected by Girvin and Jach), the much improved matching with the Coulomb
ground state's longer range oscillations is at least as striking. 
Turning to the entanglement spectrum, we find that introducing even a small nonzero $d$ brings forward the branches of the Coulomb ES at higher entanglement energies that are completely missing from the Laughlin ES.
Using the optimal values of $d$ allows for a very good qualitative fit of the entire Coulomb ES as well as a good quantitative fit at low entanglement energy. 

There can be little argument that the modified Laughlin states describe the same universality class as the usual Laughlin state.
All observables we have calculated show very smooth behavior as a function of $d$.
Of course we are limited to small system sizes, but, especially on the sphere, we appear to nevertheless reach the scaling region at least for the energy and for the optimal value of $d$ (see \figref{fig:GS_energy_scan_of_d}).
The entanglement spectra also show a stable low lying Laughlin type branch for a broad range of $d$-values. 

A natural extension of this work is a study of modified Laughlin states with excitations, such as quasiholes,
quasiparticles and excitons.
Trial wave functions for these can be constructed by applying modification factors similar to those in \eqref{eq:alt-L_basic} to the Laughlin state with excitations.
However, this is not the only possible way.
One may also introduce additional variational parameters modifying the quasihole profile or construct excitations using a CF construction based on reverse flux attachment (\eg at $d=1$).
We can also consider different filling fractions, especially $\nu=1/5$.
Early indications are that improvements over the Laughlin state similar to those at $\nu=1/3$ can be obtained there but at substantially higher values of $d$.
The energy projection can be used to study all these possibilities and we intend to report on a number of them shortly\cite{FremlingWIP}.

The fact that the energy projection is a controlled approximation allows one also to use it to test the Jain-Kamilla (JK) type projections used for numerical work on composite fermion and BS--hierarchy\cite{Bonderson08} wave functions against exact projection for larger system sizes than were possible up to now.  We are in the process of doing this as part of a larger study of reverse flux CF wave functions\cite{FulsebakkeWIP}.

The EP can also be used to evaluate the CF wave functions on the torus. Work on
these was recently done by Hermanns\cite{Hermanns13_PRB} but the wave functions
could only be evaluated for a very small number of particles. Using EP the wave
functions could be tested at larger system sizes.
We have some hope that the EP may also help alleviate computational difficulties other than the LLL-projection, notably explicit symmetrization and antisymmetrization of trial wave functions. 

Other ways to improve the Laughlin wave function include the fixed phase quantum Monte Carlo method of \refcite{Ortiz93},
which can find the optimal wave function when the phase of the function is given.
It would be interesting to compare the results from this method to the best results obtained using the single parameter family of states considered here,
and also potentially to try and further improve the modified Laughlin states using this method.
We have checked by direct analytic calculation for small systems that the phase of the modified Laughlin wave functions does depend on $d$ and in particular that it is not the same as the phase of the standard Laughlin wave function. 
Recently there has been much interest also in modifications of Hall states (including the Laughlin state) by the introduction of geometric anisotropy\cite{Haldane11,Qiu12,Yang12} and it would be interesting to generalize the modified Laughlin states to this context also. 

Going beyond the Laughlin states, modifications similar to those in \eqref{eq:alt-L_basic} can be made to any planar or spherical trial wave function.
This could thus be used to massage the CF wave functions of the Jain series,
but can also be applied to more exotic wave functions such as \eg the Moore-Read Pfaffian wave function\cite{Moore91} at $\nu=5/2$ or its generalizations such as the Read-Rezayi\cite{Read99} or BS--hierarchy\cite{Bonderson08} wave functions. 

The modification made to the Laughlin wave functions can also be easily generalized.
In fact, the modified Laughlin states are only the simplest type of modified states in a large class of wave functions which can be described using Wen's $K$-matrix formalism\cite{Wen92a}.
For any such wave function, one may split the $K$-matrix into a holomorphic and an anti-holomorphic part,
writing $K=\kappa-\bar{\kappa}$, where $\kappa$ and $\bar{\kappa}$ are both positive definite\cite{Suorsa11_PRB,Suorsa11_NewJPhys,Hansson16}.
For the Laughlin state at filling $\nu=\frac{1}{3}$, we simply have the $1\times 1$ matrix $K=3$,
with the modified states obtained using $\kappa=3+d$ and $\bar{\kappa}=d$.
For multi-layer states or states based on CF constructions with multiple Landau levels,
the $K$-matrix will be higher dimensional and many more modifications become possible.
States with counterpropagating edge modes  must be realized with nonzero $\kappa$ and $\bar{\kappa}$ and in such cases the EP may be the only way to evaluate them at reasonable system sizes.
Such non-chiral states would include for example the $\nu=2/3$ state,
especially on the torus where other approximate projection methods are not available.

We would like to stress that the division of the K-matrix into holomorphic and anti-holomorphic parts does not introduce any extra (counter-propagating) edge modes.
The chiralities of the edge modes are given by the signs of eigenvalues of the full K-matrix\cite{Wen92b}, and are independent of this decomposition.
Thus we expect to get only one edge mode for all the modified Laughlin states, in the same way that we expect one chiral and one anti-chiral edge mode in the case of $\nu=2/3$.
This is also supported by the entanglement spectra in \figref{fig:ES_sphere} and \figref{fig:ES_torus}.

\section{Computations}
\label{sec:computations} 
This project entailed significant numerical computations using a mix of freely
available codes as well as codes developed in-house by the authors. 

A set of codes, christened Hammer\footnote{http://www.thphys.nuim.ie/hammer} were
developed by the authors and were used for the majority of the computations.
These codes have many notable features. They provide a diagonalisation code with
the ability to accurately resolve large numbers of eigenstates for large sparse
matrices by employing the Krylov subspace methods provided by SLEPc
\cite{Hernandez:2005:SSF} and taking advantage of large distributed memory
machines. This code can exploit many symmetries of the hamiltonian to reduce the
computational effort and provide additional quantum numbers. It includes
utilities to calculate many other quantities including entanglement spectra, 
correlation functions and Hall viscosities. There is a significant functionality
for performing Monte Carlo (MC) simulations on both the sphere and torus, with
utilities that can efficiently evaluate many trial wave-functions.  

The DiagHam package \footnote{http://nick-ux.lpa.ens.fr/diagham/wiki} is a freely available set of utilities for performing
calculations of FQH systems. This package was used for the following
computations on the sphere: initial diagonalization calculations for small systems,
real space evaluation of Fock space wave-functions
and the calculation of entanglement spectrum and correlation functions.

\vspace*{2mm}
\acknowledgments
\noindent 
We thank Hans Hansson for pointing us in the direction of the modified Laughlin wave functions and Steve Simon and Andrei Bernevig for useful discussions.
JKS thanks Nordita for its hospitality during the program on Physics of
Interfaces and Layered Structures (2015), when part of this work was done. NM
thanks the organisers of the TOPO2015 workshop held at the Institut d'Etudes
Scientifiques de Carg\`ese where part of this work and related discussions took
place.  
All authors acknowledge financial support through SFI Principal Investigator Award
12/IA/1697.
We also wish to acknowledge the SFI/HEA Irish Centre for High-End Computing (ICHEC) for the provision of computational facilities and support.

\bibliography{corrsorted}

\end{document}